\documentstyle[floats,prd,aps,amssymb,epsf]{revtex}
\newcommand{\hn}{{\mathfrak n}}
\newcommand{\ds}{d\sigma}
\newtheorem{prop}{Proposition}[section]
\newtheorem{Def}{Definition}[section]
\newtheorem{lemma}{Lemma}[section]
\newcommand{\madm}{M_{\rm ADM}}

\newcommand{\beq}{\begin{equation}} 
\newcommand{\eeq}{\end{equation}} 
\newcommand{\beqn}{\begin{eqnarray}} 
\newcommand{\eeqn}{\end{eqnarray}} 
\newcommand{\pa}{\partial}
\newcommand{\na}{\nabla}

\newcommand{\gabu}{g^{\alpha\beta}}

\newcommand{\qabu}{q^{\alpha\beta}}

\newcommand{\Tabu}{T^{\alpha\beta}}
\newcommand{\Tab}{T^\alpha{}\!_\beta}
\newcommand{\Gab}{G^\alpha{}\!_\beta}

\newcommand{\Gabu}{G^{\alpha\beta}}
\newcommand{\Rab}{R^\alpha{}\!_\beta}
\newcommand{\dSab}{dS_{\alpha\beta}}
\newcommand{\dSa}{dS_{\alpha}}

\newcommand{\dl}{\delta}

\newcommand{\dlgab}{\delta g_{\alpha\beta}}
\newcommand{\Dlgab}{\Delta g_{\alpha\beta}}

\newcommand{\Dl}{\Delta}

\newcommand{\Lie}{\mbox{\pounds}}
\newcommand{\Lag}{{\cal L}}

\newcommand{\nalam}{\mathrel{\raise0.9ex\hbox{$^\lambda$}\mkern-14mu
\lower0.0ex\hbox{$\nabla$}}}

\begin{document} 
\preprint{UWM ...}

% 
%%%%%%%%%%%%%%%%%%%%%%%%%%%%%%%%%%%%%%%%%%%%%%%%%%%%%%%%%%%%%%%%%%%%%% 
% uncomment the following two lines and one below for two columns! 
%%%%%%%%%%%%%%%%%%%%%%%%%%%%%%%%%%%%%%%%%%%%%%%%%%%%%%%%%%%%%%%%%%%%%% 
%%\twocolumn[\hsize\textwidth\columnwidth\hsize\csname 
%%@twocolumnfalse\endcsname 
% 
 
\title{ 
Thermodynamics of binary black holes and neutron stars  
} 
 
\author{John L. Friedman$^1$ K\=oji Ury\=u$^1$ and Masaru Shibata$^2$} 
\address{ 
$^1$ 
Department of Physics, University of Wisconsin-Milwaukee, P.O. Box 413,  
Milwaukee, WI 53201, U.S.A., \\ 
$^2$ 
Department of Earth Science and Astronomy, Graduate School of Arts and  
Sciences, University of Tokyo, Komaba,  
Meguro, Tokyo 153-8902, Japan} 
% 

% 
% 
% 
%\date{\today}  
\maketitle  

\begin{abstract} 
We consider compact binary systems, modeled in general relativity as 
vacuum or perfect-fluid spacetimes with a helical Killing vector 
$k^\alpha$, heuristically, the generator of time-translations in a 
corotating frame.  Systems that are stationary in this sense are not 
asymptotically flat, but have asymptotic behavior corresponding to 
equal amounts of ingoing and outgoing radiation.  For black-hole 
binaries, a rigidity theorem implies that the Killing vector lies along 
the horizon's generators, and from this one can deduce the zeroth law 
(constant surface gravity of the horizon).  Remarkably, although the  
mass and angular momentum of such a system are not defined, there is  
an exact first law, relating the change in the asymptotic Noether charge  
to the changes in the vorticity, baryon mass, and entropy of the fluid,
and in the area of black holes. \\
  
Binary systems with $M\Omega$ small have an approximate asymptopia in
which one can write the first law in terms of the asymptotic mass and
angular momentum.  Asymptotic flatness is precise in two classes of
solutions used to model binary systems: spacetimes satisfying the
post-Newtonian equations, and solutions to a modified set of field
equations that have a spatially conformally flat metric. (The spatial
conformal flatness formalism with helical symmetry, however, is
consistent with maximal slicing only if replaces the extrinsic
curvature in the field equations by an artificially tracefree
expression in terms of the shift vector.)  For these spacetimes,
nearby equilibria whose stars have the same vorticity obey the relation
$ \delta M = \Omega \delta J,$  from which one can obtain a turning
point criterion that governs the stability of orbits.
\end{abstract} 
%\maketitle 

\section{Introduction} 
\label{intro} 
Beginning with papers by Blackburn and Detweiler \cite{bd92,d94}, 
several authors have used spacetimes with a helical Killing vector 
\footnote{In a spacetime with a rotational Killing vector $\phi^\alpha$  
and a timelike Killing vector $t^\alpha$, each combination  
$t^\alpha + \Omega \phi^\alpha$, with $\Omega$ constant and nonzero, will  
be called a helical (or helicoidal) Killing vector.  We give a precise  
definition in \ref{sec:2} and discuss its relation to a previous definition  
by Bonazzola et al \cite{bgm97}}   
to model binary systems in the context of general relativity. Such  
spacetimes can be regarded as having equal amounts of incoming and  
outgoing radiation; they are a counterpart in general relativity of  
the stationary solution due to Schild \cite{schild} that  
describes two oppositely charged particles whose electromagnetic field 
is the half-advanced + half-retarded solution of the orbiting charges. 
Because the radiation field of such a stationary solution has infinite 
energy, spacetimes that describe the corresponding general relativistic 
binaries are not asymptotically flat.  Instead, the asymptotic mass 
rises linearly with a naturally defined radial coordinate. 
 
The formal lack of asymptotic flatness has been handled in several 
related ways.  As Detweiler has emphasized, one can define an 
approximate asymptotic region for systems in which the energy emitted 
in gravitational waves in a dynamical time is small compared 
to the mass of the system.  In this ``local wave zone,'' the geometry 
describes gravitational waves propagating on a Schwarzschild 
background.  In the more restrictive context of the post-Newtonian 
approximation, one regains asymptotic flatness, because there 
is no radiation through second post-Newtonian order.  Finally, a number 
of authors \cite{wm956,ba97,bgm989,mmw99,use00} considered spacetimes with 
conformally flat spacelike slices that satisfy a truncated set of field 
equations consisting of the initial value  equations, a field equation 
for the spatial conformal factor, and the equation of hydrostatic 
equilibrium.  Like the post-Newtonian spacetimes, these conformally 
flat spacetimes are nonradiative and (as we show) are asymptotically flat.  
 
We consider binary systems modeled in the exact theory (without
asymptotic flatness) and then apply our results to the post-Newtonian
spacetimes and spatially conformally flat spacetimes that retain
asymptotic flatness.  In each case one uses the Killing vector, $k^\alpha,$
to define a conserved current and associated charges.  For the exact
vacuum and perfect-fluid spacetimes, the Noether current of the helical
Killing vector assigns to each spacetime a charge
$Q$. (See, for example, Refs. \cite{wald,iyerwald,iyer,ss77,sorkin91,brown}.)
Despite the lack of asymptotic flatness one
can choose the current to make $Q$ finite, and it $Q$ is independent of the
2-surface $S$ on which it is evaluated, as long as $S$ lies outside the 
matter and all black holes.  The Noether current assigns to
each black hole a charge that can be identified with its entropy (its
area, in the spacetimes we consider); and we obtain a version of the
first law (Eq. (\ref{eq:dQ}) below) that expresses the change $\delta Q$ 
in terms of changes in the vorticity, baryon mass, and entropy of the fluid, 
and in the area of black holes.  Independent work by Baker and Detweiler
\cite{bd01} obtains a similar first law for spacetimes with approximate 
asymptotic flatness at finite distance from the binary.

In the asymptotically flat spacetimes mentioned above, the helical Killing vector 
has the asymptotic form $k^\alpha = t^\alpha + \Omega \phi^\alpha$, where 
$t^\alpha$ and $\phi^\alpha$  generate asymptotic symmetries associated with 
time-translations and rotations.  Neighboring perfect-fluid  
equilibria in a post-Newtonian or a spatially conformally flat framework
satisfy a first law of thermodynamics of the form 
\[
	\dl M    
	= \Omega \delta J + \int_\Sigma \left[\bar{T}\Dl dS + \bar{\mu}\Dl
dM_{\rm B} 
	  + v^\alpha\Dl dC_\alpha\right] 
	  + \sum_i \frac{1}{8\pi}\kappa_i \dl A_i.   
\]
Here $M$ and $J$ are the ADM mass and angular momentum of the spacetime
(see Eqs.~(\ref{eq:madm},\ref{eq:j});
$\bar T$ and $\bar \mu$ are the redshifted temperature and chemical 
potential; $dM_B$ is the baryon mass of a fluid element; and $dC_\alpha$ is 
related to the circulation of a fluid element 
(see Eqs.~(\ref{eq:tmudef},\ref{eq:dmscdef})). 

Note that, in the full theory, models of binaries with a helical
Killing vector can only have corotating black holes.  If their 
generators do not lie along the Killing vector the black holes
will have nonzero shear and thus (assuming positive energy)
increasing area; and this is inconsistent with the assumption 
of a helical Killing vector.  In an appendix, we derive  
a virial relation for binary neutron-star systems in a conformally flat
framework and show that the relation is equivalent to the 
equality of the Komar and ADM mass.         

One other class of asymptotically flat spacetimes with a single 
Killing vector is worth mentioning.  These are nonaxisymmetric 
stars whose figure is {\it stationary in an inertial frame},
the analog in general relativity of the Newtonian Dedekind ellipsoids.
We expect that such stationary, nonaxisymmetric perfect-fluid spacetimes 
exist; their velocity fields have nonzero shear, however, and cannot 
be stationary when viscosity is present.\cite{lindblom}  \\   

Conventions:  Spacetime indices are Greek, spatial indices 
Latin, and the metric signature is $-+++$. Readers familiar with abstract 
indices can regard indices early in the alphabet as abstract, while 
$i, j, k, l$ are concrete, associated with a chart $\{x^i\}$. 
We use the dual form of Stokes' theorem for the divergence of an
antisymmetric tensor $A^{\alpha\cdots\beta\gamma}$, namely
\[
\int_\Sigma \nabla_\gamma A^{\alpha\cdots\beta\gamma}dS_{\alpha\cdots\beta}
= \int_{\partial \Sigma} A^{\alpha\cdots\beta\gamma} dS_{\alpha\cdots\beta\gamma} 
\] 
where $dS_\alpha = \epsilon_{\alpha\beta\gamma\delta} dS^{\beta\gamma\delta}$, 
$dS_{\alpha\beta} = \epsilon_{\alpha\beta\gamma\delta} dS^{\gamma\delta}$.  
For example, in an oriented chart $t, \{x^i\}$ with $\Sigma$ a surface 
of constant $t$ and $\partial\Sigma$ a surface of constant $t$ and $x^1$, 
$dS_\alpha = \nabla_\alpha t\sqrt{-g}d^3x$, 
$\displaystyle
dS_{\alpha\beta} =  \frac12(\nabla_\alpha t\nabla_\beta x^1 
		- \nabla_\alpha x^1\nabla_\beta t)\sqrt{-g}d^2x$.  Finally,
if $S$ is a 2-surface in a 3-space $\Sigma$ and $\epsilon_{abc}$ is the 
volume form on $\Sigma$ associated with a 3-metric $\gamma_{ab}$, we write 
$dS_a = \epsilon_{abc}dS^{bc}$; for $S$ a surface of constant $r$,
$dS_a = \na_a r \sqrt{\gamma} d^2 x$.

\section{{Helical Killing Vectors, Event Horizons, and the Zeroth Law}} 
\label{sec:2} 
 
We consider globally hyperbolic spacetimes
$M,g_{\alpha\beta}$ that have a  symmetry vector $k^\alpha$, a
Killing vector that generates a symmetry  of the matter
fields.  Our particular interest is in stationary binary systems, systems
whose Killing vector $k^\alpha$ has helical  integral curves with a
fixed period $T$; but our results hold for a broader class of spacetimes
with a single Killing vector.  

We begin by using the periodic orbits just mentioned to define a {\sl
helical} vector field. We want a definition that agrees, for
stationary, axisymmetric spacetimes, with Killing vectors of the form
$t^\alpha + \Omega \phi^\alpha$,   where $t^\alpha$ is the
asymptotically timelike Killing vector and $\phi^\alpha$ the 
rotational Killing vector. Let $\chi_t$ be the   family of diffeos
generated by
$k^\alpha$, moving each point $P\in M$  a parameter distance $t$ along
the integral curve of $k^\alpha$ through $P$.     Although a helical
vector is spacelike at distances from the  axis larger than
$T/{2\pi}$, its integral curves spiral each period to  points that are
timelike separated from their starting points; at least  they are
timelike separated when one is outside a finite region that  encloses
any horizon or ergosphere.  Without this last caveat, one  could
define a helical vector by the requirement that, for each
point $P$, $\chi_T(P)$ be timelike separated from $P$.  To include the
caveat, one requires that the condition hold only outside some sphere.
Let ${\cal S}$ be a spacelike sphere, and let ${\cal T}$ be the
timelike surface swept out by the action of $\chi_t$ on ${\cal  S}$:
${\cal T}({\cal S}) = \cup_t \chi_t({\cal S})$; we call ${\cal T}$
the history of ${\cal S}$.
 
\begin{Def} A vector field $k^\alpha$ is {\sl helical} if there 
is a smallest $T>0$ for which $P$ and 
$\chi_T(P)$ are timelike separated for every $P$ outside the history 
${\cal T}$ of some sphere.   
\end{Def} 
 
When the spacetime admits a foliation by timelike lines, this definition is 
equivalent to the following definition,  
essentially that of Bonazzola et al.\cite{bgm97}: 
 
\begin{prop} A vector field $k^\alpha$ is helical if it can be written in 
the form
\beq
k^\alpha = t^\alpha +\Omega \phi^\alpha, 
\label{eq:helical2}\eeq
where  $\phi^\alpha$ is spacelike and has circular orbits with
parameter  length $2\pi$, except where it vanishes; $\Omega$ is a
constant; and,  $t^\alpha$ is timelike outside the history ${\cal T}$
of some sphere. Conversely, if a vector field is helical, and if 
the spacetime can be foliated by timelike curves that respect the 
action $\chi_T(P)$, then $k^\alpha$ can be written in the form 
(\ref{eq:helical2}).
\footnote{Without the requirement on the timelike character  of
$t^\alpha$, any Killing vector can be written in the form $t^\alpha  +
\Omega\phi^\alpha$.  To restrict {\em helical} to the vector fields in
which we are interested, we had to exclude the spiral
Killing vectors of Minkowski space that have the  form $s^\alpha +
\phi^\alpha$, with $s^\alpha$ a constant {\em spacelike}  vector.}
\end{prop} 
 
Because there are spacetimes with helical vectors that do not 
allow foliations respecting the action $\chi_T(P)$, the Bonazzola et al. 
definition is slightly more restrictive than ours; they are also more 
restrictive in requiring the existence of a 2-dimensional submanifold, 
the axis of symmetry, on which $\phi^\alpha$ vanishes; and in requiring 
that $t^\alpha$ be timelike everywhere. 
Note that, although the proposition displays the intuitive character of  
a helical vector, $t^\alpha$ and $\phi^\alpha$ are far from 
unique.  Each foliation of $M$ by a family of timelike curves that  
respects $\chi_T$ gives a different decomposition of $k^\alpha$ of  
the form (\ref{eq:helical2}).\\      
 
\noindent{\sl Proof of Proposition.}    
The first part of the Proposition, that a vector of the form 
(\ref{eq:helical2}) is helical, is immediate.  We prove as follows that  
a helical vector can be written in this form.  Define a scalar by
requiring it to have the value $t$ on
$\chi_t(\Sigma)$, with $\Sigma$ a Cauchy surface. Let 
$t^\alpha$ be the vector tangent to our congruence of timelike curves, 
each parametrized by $t$.
 
Let $\psi_t$ be the family of diffeos generated by $t^\alpha$.   
Each integral curve of $k^\alpha$ can be projected to a circle  
on $\Sigma$ by pushing it down to $\Sigma$ along the timelike  
congruence: The circle through each point $P\in\Sigma$ is  
$$  
        t\rightarrow c(t) := \psi_{-t}\circ \chi_t(P). 
$$ 
One obtains a circle with parameter length $2\pi$ by reparametrizing 
$c$, defining $C(s) := c[ sT/(2\pi)]$.  Finally, define $\phi^\alpha$ 
on $\Sigma$ as the vector field tangent at each point $P$ to the  
circles $C(t)$ through $P$; and drag $\phi^\alpha$ by $\psi_t$ to 
extend it to $M$.  Then $k^\alpha = t^\alpha +\Omega\phi^\alpha$, 
with $\Omega = 2\pi/T$. 
\hfill$\Box$  
 
\vspace{2mm} 
 
In particular a spacetime that is stationary and axisymmetric, with 
asymptotically timelike Killing vector $t^\alpha$ and rotational 
Killing vector $\phi^\alpha$, has a family of helical Killing vectors 
$t^\alpha + \Omega \phi^\alpha$, for each $\Omega$.  Our primary 
concern, of course, is with binary systems, spacetimes for which 
$t^\alpha$ and $\phi^\alpha$ are not themselves Killing vectors, 
although, for one value of $\Omega$, $k^\alpha = t^\alpha + 
\Omega \phi^\alpha$ is. 
 
We have emphasized that spacetimes with a helical Killing vector cannot
be asymptotically flat in the exact theory, and a theorem by Gibbons
and Stewart\cite{gs}, showing that $\cal I$ (null infinity)
cannot be periodic, makes this
claim precise: No spacetime can have a $\cal I$ (and hence no spacetime can
be asymptotically flat) if it is vacuum outside a compact region and
has a helical Killing vector.  We can, however, use the Killing vector
$k^\alpha$ to define as follows the future and past horizon and the
future and past domains of outer communication of a spacetime with a
helical Killing vector.
 
\begin{Def}   
A point $x\in M$ is in the future (past) domain of outer 
communication, ${\cal D}^\pm$ if some future-directed (past-directed)  
timelike curve $c(\lambda)$ through $P$ eventually exits and  
remains outside the history ${\cal T}$ of each sphere $S$:   
That is, for each history ${\cal T}$ that encloses $P$, there is some  
$\lambda_0$ for which $c(\lambda)$ is outside of ${\cal T}$, all 
$\lambda>\lambda_0$.   
\end{Def} 
 
\begin{Def} The future (past) event horizon ${\cal H}^\pm$ is the boundary
of the 
future (past) domain of outer communication.   
\end{Def} 
 
\begin{prop} Let the history ${\cal T}$ of a spacelike sphere lie  
in ${\cal D}^\pm$.  Then ${\cal H}^\pm = \partial I^\mp({\cal T})$. 
\end{prop} 

\noindent 
{\sl Proof}.  Denote by int(${\cal T})$, the points inside a history  
${\cal T}$.  It suffices to show that 
$I^-({\cal D}^+)\bigcap {\rm int}({\cal T})  
= I^-({\cal T}^+)\bigcap {\rm int}({\cal T})$.  For any  
$P\in I^-({\cal D}^+)\bigcap {\rm int}({\cal T}),$ 
there is a timelike curve from $P$ that exits ${\cal T}$ and hence  
intersects ${\cal T}$.  Thus  
$P\in I^-({\cal T}^+)\cap {\rm int}({\cal T})$;  
and, from  
${\cal T}\subset {\cal D}^+ \Rightarrow I^-({\cal T})\subset I^-({\cal D}),$ 
the result follows. \hfill $\Box$ 

The main result of this section is that $\cal H^\pm$ are Killing 
horizons and hence that they satisfy the zeroth law of black-hole 
thermodynamics:   That is, the Killing vector $k^\alpha$ is tangent
on ${\cal H}^\pm$ to the null generators;  and the associated 
surface gravity $\kappa$, defined by   
\beq 
	k^\beta\nabla_\beta k^\alpha = \kappa k^\alpha,
\label{eq:kappa}
\eeq
is constant on each connected component of ${\cal H}^\pm$. \\ 

To prove that ${\cal H}^\pm$ is a Killing horizon
(Prop. \ref{prop:kill} below), we will use an analogous theorem proved
by Isenberg and Moncrief  \cite{mi83,im85} and a strengthened version by
Friedrich, R\'acz, and Wald\cite{frw} (henceforth FR\&W), for a class
of spacetimes with a compact null surface (see also earlier work by 
Hawking\cite{hawking72}).  Following FR\&W, we first
show that the spacetime  $N,g_{\alpha\beta}$ covers such a compact
spacetime.  Although our spacetime is not in the class they study, FR\&W
note that their asymptotic conditions can be relaxed, and
we easily extend their proof to spacetimes of the kind considered
here.
 
For convenience in matching our definition and proof to that of FR\&W,  
we consider a subspacetime $N = I^+(M)\bigcap \overline{ {\rm int} {\cal T}
}$,  
for some ${\cal T}$ that encloses the fluid and black holes.  
By choosing a future set, we keep all black holes but discard the
bifurcation horizon and white holes that are part of the full
spacetime.  (To obtain the corresponding results for white holes -- for
the past horizon -- one exchanges future and past). When the surface 
gravity $\kappa$ is nonzero, the past-directed null generators reach the 
bifurcation horizon of $M$ in finite affine parameter length.  This means 
that in $N$, they are past geodesically incomplete, and that past 
incompleteness is one of the conditions required for the FR\&W proof.
The Isenberg-Moncrief version does not require past incompleteness, 
but does demand that the horizon be analytic.  $N,g_{\alpha\beta}$ satisfies 
the following conditions that define a spacetime of type A$'$.

\begin{Def} A smooth spacetime $N,g_{\alpha\beta}$ will be said to be of 
class A$'$ if it has the following properties.\\ 
(i)  The spacetime has a Killing vector field $k^\alpha$ that  
is transverse to a Cauchy surface.\footnote{A vector field $k^\alpha$ 
is {\em transverse} to a hypersurface $S$ if $k^\alpha$ is nowhere zero
on $S$ and nowhere tangent to $S$.}  
\\ 
(ii) $N = I^+(N)$ \\
(iii) There is a history ${\cal T}$ for which $N = I^+({\cal T})$\\
(iv) The horizon ${\cal H} :=\partial I^-(N)$ consists of smooth
disconnected  
components each of which has topology ${\Bbb R}\times S^2$. \\
(v) The generators of ${\cal H}$ are past incomplete. (Alternatively,
   ${\cal H}$ is analytic.)
\end{Def}

\begin{prop} Let $N,g_{\alpha\beta}$ be a spacetime of type A$'$, satisfying 
the null energy condition $R^{\alpha\beta}l^\alpha l^\beta \geq 0,$ all 
null $l^\alpha$.  Then on each component of the horizon, there exists a  
$t_0 \neq 0$ such that $\chi_{t_0}$ maps each null geodesic generator of  
$\cal H$ to itself. 
\label{prop:gen}\end{prop}

We first need to establish for spacetimes of class A$'$ an analog of 
Prop. 9.3.1 of Hawking and Ellis\cite{he73},  showing that the shear 
and divergence of the horizon generators vanish. This implies that 
the generators are Killing vectors of the horizon, Lie-deriving 
its degenerate 3-metric.

\begin{lemma} Let $N,g_{\alpha\beta}$ be a spacetime of class A$'$.  
On each component of the horizon,  the shear and expansion of the null 
generators vanishes. 
\label{lemma}\end{lemma} 
 
\noindent{\sl Proof of Lemma \ref{lemma}}. Let $S$ be a Cauchy surface
transverse to $k^\alpha$, $S_t = \chi_t(S)$, and let ${\cal B}_t =
S_t\bigcap {\cal H}$.  Because $\chi_t$ is an isometry, it maps
${\cal H}$ to itself. Then $k^\alpha$ is tangent to  ${\cal H}$, and
the family of slices ${\cal F}_t$ foliates ${\cal H}$.  Because ${\cal
F}_t$ is mapped to ${\cal F}_{t'}$ by the isometry $\chi_{t'-t}$, the
area of ${\cal F}_t$ is independent of $t$.  This implies that the
divergence $\theta$ of the horizon's generators vanishes and that the
generators have no past endpoints.  Finally, using $\theta=0$ and the
null energy condition, the Raychaudhuri equation (optical scalar
equation), 

\beq   
\frac{d\theta}{d\lambda}   =
	- R_{\alpha\beta}l^\alpha l^\beta 
	- 2\sigma_{\alpha\beta}\sigma^{\alpha\beta}-\frac12 \theta^2,  
\eeq
implies $\sigma_{\alpha\beta}=0$.   \hfill$\Box $
 
\noindent{\sl Proof of Proposition \ref{prop:gen}}.  Once Lemma
\ref{lemma} is proved, the  proof of this proposition is exactly the
proof of Prop. 2.1 in  FR\&W.\hfill$\Box$

\begin{Def} A spacetime $N, g_{\alpha\beta}$ is of class B if it contains
a compact orientable, smooth null hypersurface ${\cal N}$ that is generated
by closed null geodesics.  
\end{Def}
(These causality-violating spacetimes are 
introduced only as part of the proof of Prop. \ref{prop:ab}; 
the spacetimes considered in this paper  
as models of physical systems are globally hyperbolic.) 

\begin{prop} Let $N,g_{\alpha\beta}$ be a spacetime of type A$'$. Then 
(int$(N), g_{\alpha\beta})$ is a covering spacetime of a spacetime of type B.
\label{prop:ab}\end{prop}

The proof is immediate:\\
\noindent{\sl Proof of Prop. \ref{prop:ab} }.  Because $k^\alpha$ is
transverse to a Cauchy surface, $\chi_t$ has no fixed points for
$t\neq 0$; in particular, for $t = t_0$  of Prop. \ref{prop:gen},
$\chi_{t_0}$ has no fixed points.  Then the factor space $\tilde N =
{\rm int}(N)/\chi_{t_0}$, with induced metric  $\tilde
g_{\alpha\beta}$ has covering spacetime ${\rm int}(N),g_{\alpha\beta}$.
Because $\chi_{t_0}$  maps each generator of ${\cal H}$ to itself,
$\tilde{\cal H} = {\cal H}/\chi_{t_0}$ is a null hypersurface
generated  by closed null geodesics.  \hfill$\Box$\\

\begin{prop}  In a spacetime of class A$'$ ${\cal H}$ is a Killing horizon 
of $k^\alpha$.  In particular, if, up to a constant scaling, $k^\alpha$ is
the 
only Killing vector in $N$ (or in any subspacetime), then $k^\alpha$ is 
parallel to the null generators of $\cal H$.  
\label{prop:kill}\end{prop}
  
\noindent{\sl Proof}.  Any neighborhood of a component of the horizon of 
(int$(N), g_{\alpha\beta})$ that is disjoint from the fluid 
covers a vacuum spacetime of type B.  Theorem 4.1 of FR\&W implies that 
in a one-sided neighborhood of that component of the horizon, there is a 
Killing vector $\tilde K^\alpha$ normal to the the horizon.  The pullback 
$K^\alpha$ of $\tilde K^\alpha$ to the covering space is then a Killing
vector 
on a one-sided neighborhood of the corresponding component of $\cal H$, 
normal on $\cal H$ to $\cal H$: i.e., $\cal H$ is a Killing horizon.  
If each neighborhood has $k^\alpha$ as its only Killing vector (up to an 
overall scale),then $k^\alpha \propto K^\alpha$ on each component of ${\cal
H}$,
implying that $\cal H$ is a Killing horizon with Killing vector $k^\alpha$.
\hfill $\Box$\\
 
\noindent{\bf Corollary (0th Law)}. The surface gravity $\kappa_i$ is
constant on 
the ith component of $\cal H$. \\ 

\noindent{\sl Proof}.  The proof of the zeroth law of event horizons 
given in Bardeen et al.\cite{bch73} establishes the result for any 
Killing horizon in a spacetime satisfying the null energy condition. \\    

The first law is the content of the next section. The second law, that
the area of a black hole cannot decrease, has meaning here only if one
extend the definition of event horizon in a way that requires neither a
Killing vector nor asymptotic flatness.   Black-hole thermodynamics of
general spacetimes that are not asymptotically flat has been examined
previously \cite{gh77,isolatedbh1,isolatedbh2,penrose} but the results
here appear to be new.  \footnote{In particular, in the isolated horizon
framework, for a horizon with a single Killing vector, one shows the
existence of a charge $E$ defined on an isolated horizon for which 
$\delta E = \kappa \delta A$;\cite{isolatedbh2} in our case this 
is satisfied by the charge $\delta Q_i=\delta Q_{Li}+\delta Q_{Ki}$ 
defined on the $i$th disconnected component of the horizon by 
Eqs. (\ref{46}) and (\ref{50}). Our first law, in contrast, relates 
this change in the black-hole charges to the changes in the Noether 
charge of a sphere surrounding all black holes and all matter and 
to the changes in the entropy, baryon number, and circulation of the fluid.
The existence of such a first law depends precisely on what is {\it not} 
assumed in the isolated horizon framework: a globally defined Killing 
vector.}
  
\section{First law for systems with a single Killing vector} 
\label{sec:3} 
 
We consider spacetimes with black holes and perfect-fluid sources,
which have a helical Killing vector or, more generally, a single 
Killing vector that is transverse to a Cauchy surface and timelike on the
support of the fluid.  Although such spacetimes will not, in general,
be asymptotically flat, one can obtain a generalized first law of
thermodynamics in terms of a Noether charge $Q$ associated with the
Killing vector field and with an action for the perfect-fluid
spacetime.  For spacetimes that are asymptotically flat, the overall
scaling of a timelike Killing vector is chosen by requiring it to have
unit norm at spatial infinity.  Here, without asymptotic flatness, the
overall scaling cannot be so determined.  Instead, in our discussion of
the first law, the choice of a family of spacetimes will include the
choice of a Killing vector; but readers should keep in mind that
nothing in this section restricts the freedom to choose another scaling
of the Killing vector for each member of the family of spacetimes.  \\

	We describe a perfect fluid by its four-velocity $u^\alpha$ and  
stress tensor  
\beq 
	T^{\alpha\beta} = \epsilon u^\alpha u^\beta + p q^{\alpha\beta}, 
\eeq	 
where $p$ is the fluid's pressure, $\epsilon$ its energy density, and  
\beq  
	q^{\alpha\beta} = g^{\alpha\beta} + u^\alpha u^\beta 
\eeq 
is the projection orthogonal to $u^\alpha$.  We assume that the fluid  
satisfies an equation of state of the form  
\beq  
	 p = p(\rho,s),\    \epsilon = \epsilon(\rho,s),   
\eeq   
with $\rho$ the baryon-mass density and $s$ the entropy per unit baryon
mass.   
(That is, $\rho := m_B n$, with $n$ the number density of baryons  
and $m_B$ the average baryon mass.)   
 
Given a family of perfect-fluid spacetimes specified by  
\beq 
{\cal Q}(\lambda) := [g_{\alpha\beta}(\lambda), u^\alpha(\lambda),  
\rho(\lambda), s(\lambda)],   
\eeq 
one defines the eulerian change in each quantity by  
$\delta {\cal Q} := \frac d{d\lambda} {\cal Q}(\lambda).$ 
 
We introduce a lagrangian displacement  
$\xi^\alpha$ in the following way: Let ${\cal Q} := {\cal Q}(\lambda)$, and let  
$\Psi_\lambda$ be a diffeo mapping each trajectory (worldline) of the
initial  
fluid to a corresponding trajectory of the configuration ${\cal Q}(\lambda)$. 
Then the tangent $\xi^\alpha(P)$ to the path  
$\lambda \rightarrow \Psi_\lambda(P)$ can be regarded as a vector 
joining the fluid element at $P$ in the configuration ${\cal Q}(\lambda)$ to a  
fluid element in a nearby configuration. The lagrangian change 
in a quantity at $\lambda=0$,\footnote{The lagrangian change is analogously 
defined at any $\lambda_0$:  The diffeo 
$\widetilde\Psi_\lambda = \Psi_{\lambda+\lambda_0} \Psi^{-1}_{\lambda_0}$
maps each fluid trajectory in the configuration ${\cal Q}(\lambda_0)$  to 
the corresponding trajectory of ${\cal Q}(\lambda+\lambda_0)$, whence
$\Delta  {\cal Q}(\lambda_0) := \frac d{d\lambda} 
	\widetilde\Psi_{-\lambda}{\cal Q}(\lambda+\lambda_0)|_{\lambda = 0}$.} 
is then given by  
\beq 
	\Delta  {\cal Q} := \frac d{d\lambda} \Psi_{-\lambda}{\cal Q} 
	(\lambda)|_{\lambda = 0}   
	  = (\delta + \Lie_\xi ){\cal Q}. 
\eeq 
 
The first law will be written in terms of integrals over a spacelike  
hypersurface $\Sigma$, transverse to $k^\alpha$, whose boundary  
\beq 
	\pa \Sigma = S \cup_i {\cal B}_i , 
\eeq 
is the union of black hole boundaries ${\cal B}_i$ (${\cal B}_i$, is  
the ith disconnected component of $\Sigma \cap {\cal H}^+$),  
and a 2-sphere $S$ that encloses  
the fluid and all black holes.  Define a scalar $t$ by setting $t=0$ on 
$\Sigma$ and requiring $k^\alpha \na_\alpha t=1$.

We can write $u^\alpha$ in the form,  
\beq 
	u^\alpha = u^t(k^\alpha+v^\alpha),  
\label{eq:udec}\eeq 
with $u^t : = u^\alpha \nabla_\alpha t$ and $v^\alpha$ a vector field on 
$\Sigma$,  
\beq 
	v^\alpha\na_\alpha t = 0.  
\eeq 
The fact that $\Psi_\lambda$ maps fluid trajectories to fluid trajectories  
and the normalization $u^\alpha u_\alpha=-1$ imply \cite{jf,carter,taub}  
\beq 
	\Dl u^\alpha = \frac{1}{2}u^\alpha u^\beta u^\gamma 
	\Delta g_{\beta\gamma}.
\label{du}   
\eeq

	One obtains an action for a perfect-fluid spacetime by considering
perturbations for which the entropy and baryon mass of each fluid element 
are conserved; and we use this action to define a Noether charge 
$Q$ associated with $k^\alpha$, for each spacetime ${\cal Q}(\lambda)$.  
Then for general perturbations, 
in which the entropy and baryon mass of each fluid element are unconstrained,
we use the charge $Q$ to write a form of the first law for perfect-fluid 
spacetimes that have one Killing vector and a Killing horizon (and that 
are not, in general, asymptotically flat).

 When the entropy and baryon mass of each fluid element are conserved along 
the family ${\cal Q}(\lambda)$, we have  
\beq 
	\Dl s = 0 \qquad {\rm and } \qquad  
	\Dl({\rho u^\alpha \sqrt{-g}})=0,  
\eeq 
implying  
\beq 
	\frac{\Dl\rho}{\rho} = - \frac{1}{2}\qabu\Dlgab;  
\label{drho}\eeq 
and the local first law of thermodynamics for the fluid,  
\beq 
	\Dl\epsilon = \rho T \Dl s + h \Dl\rho,  
\eeq 
with  
\beq 
	h = \frac{\epsilon+p}{\rho},  
\eeq 
yields  
\beq 
	\frac{\Dl\epsilon}{\epsilon+p} = \frac{\Dl\rho}{\rho} 
	=-\frac{1}{2}\qabu\Dlgab.  
\eeq 
  
  From these relations, it follows that the scalar density  
\beq 
	\Lag = \left(\frac{1}{16\pi}R - \epsilon \right)\sqrt{-g} 
\label{lag1}
\eeq 
is a lagrangian density for a perfect fluid space time.  That is,  
\beq 
	\dl\Lag = \frac{1}{16\pi}\dl(R \sqrt{-g})  
	- \Dl(\epsilon\sqrt{-g})  
	+ \na_\alpha(\epsilon\xi^\alpha)\sqrt{-g} , 
\eeq 
and (when $\Dl s = 0$ and $\Dl(\rho u^\alpha \sqrt{-g})=0$), we have,  
\beqn 
	\frac{1}{\sqrt{-g}} \Dl(\epsilon\sqrt{-g})  
	&=& - \frac{1}{2} (\epsilon+p)\qabu\Dlgab 
	  + \frac{1}{2} \epsilon\gabu\Dlgab 
\nonumber \\ 
	&=& - \frac{1}{2} \Tabu\Dlgab 
\nonumber \\ 
	&=& - \frac{1}{2} \Tabu\dlgab + \xi_\alpha\na_\beta\Tabu 
	  - \na_\alpha(\Tabu\xi_\beta).   
\label{de}\eeqn 
That ${\cal L}$ is a lagrangian density is then expressed by the equation 
\cite{WALD}
\beq 
	\frac{1}{\sqrt{-g}}\dl\Lag  
	= - \frac{1}{16\pi}(\Gabu-8\pi\Tabu)\dlgab 
	  - \xi_\alpha\na_\beta\Tabu + \na_\alpha\Theta^\alpha,  
\eeq 
with  
\beq 
	\Theta^\alpha 
	= (\epsilon+p)\qabu\xi_\beta 
	+ \frac{1}{16\pi}(g^{\alpha\gamma}g^{\beta\delta} 
	                 -g^{\alpha\beta}g^{\gamma\delta}) 
	  \na_\beta\dl g_{\gamma\delta}.   
\eeq 

Now one can associate with $\Lag$ a family of Noether charges 
\cite{ss77,wald,iyerwald,iyer,sorkin91,brown}
\footnote{Our Noether formalism is similar to Iyer's extension of 
the Wald-Iyer work to perfect fluid spacetimes.\cite{iyer}.  
Like Schutz and Sorkin,\cite{ss77,sorkin91}, 
however, we use vectors instead of forms, and our lagrangian 
displacement arises from a map $\Psi_\lambda$ from the manifold 
to itself, not, as in Iyer, from a projection onto the manifold of fluid 
trajectories.}  
\beq 
	Q = \oint_S Q^{\alpha\beta}\dSab,
\eeq
where 
\beq 
	Q^{\alpha\beta} 
	= - \frac{1}{8\pi}\na^\alpha k^\beta 
	  + k^\alpha B^\beta - k^\beta B^\alpha,  
\eeq 
and $B^\alpha(\lambda)$ is any family of vector fields that satisfies  
\beq 
    \frac1{\sqrt{-g}}\frac d{d\lambda}( B^\alpha\sqrt{-g})  = \Theta^\alpha.  
\eeq 
By choosing $B^\alpha(0) = 0$, we make $Q(\lambda)$ finite; and, 
as we will see, $Q$ is
{\it independent of the sphere $S$}, as long as $S$ encloses the fluid
and any black holes. Outside the matter,   
\beq
	B^\alpha 
	= \frac1{16\pi}(g^{\alpha\gamma}g^{\beta\delta} 
	                 -g^{\alpha\beta}g^{\gamma\delta}) 
			|_{\lambda=0}  
	  \overcirc{\na}_\beta g_{\gamma\delta}(\lambda) +\rm O(\lambda^2),  
\label{eq:b}\eeq 
where, $\overcirc{\na}_\beta$ is the
covariant derivative of the metric $g_{\alpha\beta}(0)$.

The generalized first law will be found by evaluating  
the change $\dl Q$ in this Noether charge, allowing 
perturbations that change the baryon number and entropy of each 
fluid element.  We restrict the gauge in two ways:  We use the 
diffeomorphism gauge freedom to set $\delta k^\alpha =0$.  
The description of fluid perturbations in terms of a lagrangian 
displacement $\xi^\alpha$ has a second 
kind of gauge freedom: a class of trivial displacements, including 
all displacements of the form $f u^\alpha$, yield no eulerian 
change in the fluid variables.  We use this freedom to 
set $\Delta t =0$.  Because $\delta t =0$ ($t$ is not dynamical), 
this is equivalent to the condition $\xi^t = 0$.  
Eq. (\ref{du}) now implies    
\beq
\frac{\Delta u^t}{u^t} = \frac12 u^\alpha u^\beta 
	\Delta g_{\alpha\beta}. 
\label{dgamma}
\eeq
Then, from Eqs. (\ref{du}) and (\ref{dgamma}), we have 
$\Delta u^\alpha = \Delta u^t (k^\alpha + v^\alpha)$, while, by  Eq. 
(\ref{eq:udec}),  
$\Delta u^\alpha = \Delta [ u^t (k^\alpha + v^\alpha)];$
thus
\beq 
\Delta (k^\alpha + v^\alpha) = 0.
\label{dv}  
\eeq

For perturbations that include changes in baryon number and entropy,
Eqs. (\ref{drho}) and (\ref{de}) are replaced by 
\beq 
	\frac{\Dl\rho}{\rho} = - \frac{1}{2}\qabu\Dlgab 
		+ \frac{\Dl(\rho\sqrt{-g} u^t)} {\rho\sqrt{-g} u^t}, 
\label{drho1}\eeq 
and
\beq 
 \frac{1}{\sqrt{-g}} \Dl(\epsilon\sqrt{-g}) = \rho T\Delta s 
 	+\frac h{ u^t\sqrt{-g}} \Dl(\rho u^t\sqrt{-g})
  	- \frac{1}{2}\Tabu\dlgab+\xi_\alpha\na_\beta\Tabu
	-\na_\alpha(\Tabu\xi_\beta);    
\label{de1}\eeq  
and the change in the lagrangian density becomes,
\beq
	\frac1{\sqrt{-g}}\dl\Lag 
	= - \rho T\Dl s 
	    - \frac{h}{ u^t\sqrt{-g}}\Dl (\rho u^t\,\sqrt{-g})
	    - \frac{1}{16\pi}(\Gabu-8\pi\Tabu)\dlgab  
            - \xi_\alpha\na_\beta\Tabu
	    + \na_\alpha\Theta^\alpha. 
\label{dl1}\eeq

	To find the change $\delta Q$ in the Noether charge, 
we first compute the difference, 
\beq
	\delta [Q - \sum_i Q_i], 
\eeq 
between the charge on the sphere $S$ and the sum of the charges 
on the black holes ${\cal B}_i$.  As we show below, {\it this quantity 
is invariant under gauge transformations that respect the Killing symmetry}.  
Write $Q=Q_K+Q_L$ ($Q_K$ the Komar charge, $Q_L$ an additional contribution  
involving the lagrangian density), with
\beq
 Q_K = -\frac{1}{8\pi} \oint_S \nabla^\alpha
 	k^\beta dS_{\alpha\beta},\qquad
 \delta Q_L 
	= \oint_S (k^\alpha \Theta^\beta - k^\beta \Theta^\alpha
	)dS_{\alpha\beta} ,
\eeq
\beq
	\delta (Q-\sum_i Q_i) = \delta (Q_K-\sum_i Q_{Ki}) + \delta 
	( Q_L-\sum_i Q_{Li}) .
\label{eq:delQ}
\eeq
{}From the identity 
\beq
\nabla_\beta\nabla^\alpha k^\beta = \Rab k^\beta,
\label{eq:kill}
\eeq 
we have
\beqn
Q_K -\sum_i Q_{Ki} &=& -\frac{1}{8\pi} \oint_{\pa\Sigma} \nabla^\alpha
k^\beta dS_{\alpha\beta} 
   =  -\frac{1}{8\pi} \int_\Sigma \Rab k^\beta dS_\alpha \label{eq:komar}\\
&=& - \frac{1}{8\pi} \int_\Sigma G^\alpha_\beta k^\beta dS_\alpha 
	- \frac{1}{16\pi} \int_\Sigma R k^\alpha dS_\alpha. 
\label{35}\eeqn
Now  
\beqn 
	-\Tab k^\beta\dSa 
	&=& - \Tab(k^\beta+v^\beta)\dSa + \Tab v^\beta\dSa 
\nonumber \\ 
	&=& \epsilon\,k^\alpha \dSa 
	+ (\epsilon + p) u^\alpha u_\beta v^\beta \dSa,  
\eeqn 
whence 
\beq 
	Q_K - \sum_i Q_{Ki}    
	= - \int_\Sigma \left(\frac{1}{16\pi} R - \epsilon\right) 
	    k^\alpha\dSa 
	  + \int_\Sigma (\epsilon + p) u^\alpha u_\beta v^\beta \dSa 
	    - \int_\Sigma \frac{1}{8\pi}(\Gab - 8\pi\Tab)k^\beta \dSa.  
\eeq 
and
\beqn 
	\delta (Q_K -\sum_i Q_{Ki})
	&=& - \int_\Sigma \dl {\cal L}\, d^3x 
	 + \int_\Sigma \Dl\left[(\epsilon + p) u^\alpha u_\beta  
	    v^\beta \dSa\right] 
	  \nonumber \\ 
	&& 
	  - \dl \int_\Sigma \frac{1}{8\pi}(\Gab - 8\pi\Tab) 
	    k^\beta \dSa.   
\label{dqk}\eeqn 
The second term on the right of Eq.\ (\ref{eq:delQ}) is given by
\beqn 
	\delta ( Q_L -\sum_i Q_{Li})
	&=& \oint_{\partial\Sigma}(k^\alpha \Theta^\beta - k^\beta
\Theta^\alpha)\dSab 
\nonumber \\ 
	&=& \int_\Sigma\na_\beta 
	        (k^\alpha \Theta^\beta - k^\beta \Theta^\alpha)\dSa 
\nonumber \\ 
	&=& \int_\Sigma\na_\beta\Theta^\beta k^\alpha\dSa 
	- \int_\Sigma \Lie_k \Theta^\alpha \dSa,   
\label{dq2}\eeqn 
where we have used the relation $\nabla_\alpha k^\alpha =0$ 
to obtain the last equality, and
$\nabla_\beta\Theta^\beta$ is given by Eq.\ (\ref{dl1}).  Then, adding
Eqs.\ (\ref{dqk}) and (\ref{dq2}), and using the relations
\beq 
	\Dl\left[(\epsilon + p) u^\alpha u_\beta v^\beta \dSa\right] 
	=\ hu_\beta v^\beta \Dl(\rho u^\alpha\dSa)\  
	+\ v^\beta \Dl(hu_\beta)\rho u^\alpha\dSa\ 
	+\ (\epsilon + p) u^\alpha u_\beta\, \Lie_k \xi^\beta \dSa,  
\eeq 
where $\Dl v^\beta = -\Dl k^\beta = \Lie_k \xi^\beta$ is used and 
\beq 
	\Lie_k \Theta^\alpha \dSa\ 
	=\ (\epsilon + p) q^\alpha{}\!_\beta\, \Lie_k \xi^\beta \dSa\ 
	=\ (\epsilon + p) u^\alpha u_\beta\, \Lie_k \xi^\beta \dSa, 
\eeq 
where $\Lie_k \xi^\beta\, \na_\beta t= 0$ is used, 
we obtain an expression for $\delta (Q-\sum\limits_i Q_i)$: 
\beqn 
	\delta (Q-\sum_i Q_i)  
	&=& \int_\Sigma \left[\frac{T}{ u^t} \Dl s\ \rho u^\alpha \dSa 
	  + \left(\frac{h}{u^t}+hu_\beta v^\beta\right) 
	    \Dl(\rho u^\alpha\dSa) 
	  + v^\beta\Dl(hu_\beta)\rho u^\alpha\dSa\right] 
\nonumber \\ 
	&&- \frac{1}{8\pi}\dl\int_\Sigma (\Gab - 8\pi\Tab)k^\beta\dSa 
	  + \int_\Sigma \left[\frac{1}{16\pi}(\Gabu - 8\pi\Tabu)\dlgab 
	  + \xi^\beta\na_\alpha T^\alpha_\beta \right] k^\gamma dS_\gamma. 
\label{41}   
\eeqn 

We next evaluate the black-hole charges $Q_i$.  Recall that, by
Prop.~(\ref{prop:kill}), $k^\alpha$ is tangent, on each disconnected
component ${\cal H}_i$, to the null generators of the horizon, with
surface gravity $\kappa_i$ given by 
\beq
	k^\beta\nabla_\beta k^\alpha = \kappa_i k^\alpha.
\eeq
On each ${\cal B}_i$, let $\hn^\alpha$ be the unique null vector field 
orthogonal to ${\cal B}_i$ and satisfying $\hn_\alpha k^\alpha = - 1$. 
The area element of ${\cal B}_i$ is then 
\beq
	dS_{\alpha\beta} = \frac12(k_\alpha \hn_\beta - k_\beta \hn_\alpha)dA.
\eeq
Using the Killing equation, 
$\nabla^\alpha k^\beta = \nabla^{[\alpha} k^{\beta]}$, and Eq.~(\ref{eq:kappa})
to evaluate the integrand of $Q_{Ki}$, we have 
\beq
	\nabla^\alpha k^\beta\frac12(k_\alpha \hn_\beta - k_\beta \hn_\alpha) =
		k^\beta\nabla_\beta k^\alpha \hn_\alpha = -\kappa_i,
\eeq
implying
\beq
 Q_{Ki} = -\frac{1}{8\pi}\oint_{{\cal B}_i} \na^\alpha k^\beta\dSab
  = \frac{1}{8\pi}\kappa_i A_i. \label{45}
\eeq	

Finally, following Bardeen et al. \cite{bch73}, we show that
\beq
\delta Q_{Li} = -\frac{1}{8\pi} \delta \kappa_i A_i . \label{46}
\eeq
Using $\delta (\nabla_\alpha k_\beta ) = \delta (\nabla_{[\alpha}k_{\beta
]}) = \nabla_{[\alpha}\delta k_{\beta ]}$, we have
\beqn
\delta\kappa_i &=& \delta (\hn^\alpha k^\beta\nabla_\alpha k_\beta )\nonumber\\
&=& \delta \hn^\alpha k^\beta \nabla_\alpha k_\beta + \hn^\alpha k^\beta
\nabla_{[\alpha} \delta k_{\beta ]} .\label{47}
\eeqn
Because the horizon is unchanged in our gauge, and $k_\alpha$ is parallel
to the null normal to ${\cal H}_i$, $\delta k_\alpha = ak_\alpha$, some
function $a$ on ${\cal H}_i$.  Then
\beqn
\delta \hn^\alpha k^\beta\nabla_\alpha k_\beta &=& -\delta \hn^\alpha\kappa_i
k_\alpha = \kappa_i \hn^\alpha \delta k_\alpha =-\kappa_ia\nonumber\\
&=& -a\hn^\alpha k_\beta\nabla_\alpha k^\beta =-\hn^\alpha\delta
k_\beta\nabla_\alpha k^\beta\nonumber\\
&=& k^\alpha \hn^\beta\nabla_\alpha\delta k_\beta , \label{48}
\eeqn
where, in the last equality, we have used $\Lie_k\delta k_\alpha =0$.
{}From Eqs.\ (\ref{47}) and (\ref{48}) and from the vanishing of
$\delta\sigma_{\alpha\beta}$ and $\delta\theta$, we have
\beqn
\delta\kappa_i &=& \frac{1}{2} (k^\alpha \hn^\beta + k^\beta \hn^\alpha )
\nabla_\alpha\delta k_\beta =-\frac{1}{2} \nabla^\alpha\delta
k_\alpha\nonumber\\
&=& - \frac{1}{2} k^\alpha\nabla^\beta\delta g_{\alpha\beta} .\label{49}
\eeqn
Now
\beqn
\delta Q_{Li} &=& \oint_{{\cal B}_i} (k^\alpha\Theta^\beta
-k^\beta\Theta^\alpha )dS_{\alpha\beta}\nonumber\\
&=& \frac{1}{8\pi} \oint_{{\cal B}_i} k^\alpha
(g^{\beta\delta}g^{\gamma\epsilon}
-g^{\beta\gamma}g^{\delta\epsilon})\nabla_\gamma\delta g_{\delta\epsilon}\,
\frac{1}{2} (k_\alpha \hn_\beta -k_\beta \hn_\alpha )dA\nonumber \\
&=& \frac{1}{16\pi} \oint_{{\cal B}_i} k^\alpha\nabla^\beta\delta
g_{\alpha\beta}dA\nonumber\\
&=& -\frac{1}{8\pi} \delta \kappa_i A_i \label{50}
\eeqn

The first law now follows from Eq.\ (\ref{41}) for $\delta (Q-\sum\limits_i
Q_i)$, Eq.\ (\ref{45}) for $Q_{K i}$, and Eq.\ (\ref{50}) for $\tilde
Q_{Li}$:
\beqn 
	\delta Q  
	&=& \int_\Sigma \left[\rho\frac{T}{u^t} \Dl s u^\alpha \dSa 
	  + \left(\frac{h}{u^t}+hu_\beta v^\beta\right) 
	    \Dl(\rho u^\alpha\dSa) 
	  + v^\beta\Dl(hu_\beta)\rho u^\alpha\dSa\right] 
	  + \sum_i \frac{1}{8\pi}\kappa_i\delta A_i
\nonumber \\ 
	&&- \frac{1}{8\pi}\dl\int_\Sigma (\Gab - 8\pi\Tab)k^\beta\dSa 
	  + \int_\Sigma \left[\frac{1}{16\pi}(\Gabu - 8\pi\Tabu)\dlgab 
	  + \xi^\beta\na_\alpha\Tab \right] k^\gamma dS_\gamma. \label{51}   
\label{eq:dQ0}
\eeqn 
When the family of spacetimes satisfies the field equations, the last  
line vanishes and we obtain a first law of thermodynamics in the form  
\beq 
	\dl Q   
	= \int_\Sigma \left[\rho\frac{T}{u^t} \Dl s u^\alpha \dSa 
	  + \left(\frac{h}{u^t}+hu_\beta v^\beta\right) 
	    \Dl(\rho u^\alpha\dSa) 
	  + v^\beta\Dl(hu_\beta)\rho u^\alpha\dSa\right] 
	  + \sum_i \frac{1}{8\pi}\kappa_i \dl A_i. 
\label{eq:dQ}\eeq 
Equivalently, writing  
\beq 
	\bar{T} := \frac{T}{u^t}, \qquad 
	\bar{\mu} := \frac{\mu}{u^t m_{\rm B}}  
	= \frac{h-Ts}{u^t},  
\label{eq:tmudef}\eeq 
and  
\beq 
	dM_{\rm B} := \rho u^\alpha \dSa, \qquad  
	dS:= s\ dM_{\rm B}, \qquad 
	dC_\alpha:= hu_\alpha dM_{\rm B},  
\label{eq:dmscdef}\eeq 
we have  
\beq 
	\dl Q   
	= \int_\Sigma \left[\bar{T}\Dl dS + \bar{\mu}\Dl dM_{\rm B} 
	  + v^\alpha\Dl dC_\alpha\right] 
	  + \sum_i \frac{1}{8\pi}\kappa_i \dl A_i.   
\eeq 
 
The relation between this form and that for an asymptotically flat spacetime  
with two Killing vectors, $t^\alpha$ and $\phi^\alpha$, will be found in  
Sect.~\ref{sec:af}.

	We noted above that the difference $\delta (Q-Q_i)$ is gauge invariant.
In fact, we can see as follows that $\delta (Q_K-\sum_i Q_{Ki})$ and $\delta
( Q_L-\sum_i Q_{Li})$ are separately invariant under gauge
transformations that respect the symmetry $k^\alpha$.  The gauge
transformation associated with a vector field $\eta^\alpha$ is given by
\beq
\delta_\eta {\cal Q} = \Lie_\eta{\cal Q} , \hspace{7mm} \xi^\alpha (\eta
)=-\eta^\alpha . \label{56}
\eeq
The corresponding lagrangian change in any quantity is then identically zero:
\beq
	\Delta_\eta = \delta_\eta + \Lie_{-\eta} = 0 . 
\label{57}
\eeq
{}From Eq.\ (\ref{dq2}), (\ref{dl1}), and (\ref{57}) the change in $\delta
( Q_L-\sum_i Q_{Li})$ due to a gauge transformation is given by
\beq
\delta ( Q_L-\sum_i  Q_{Li}) = \int_\Sigma \nabla_\beta \Theta^\beta
k^\alpha dS_\alpha = \int_\Sigma \delta_\eta\Lag\; d^3x , \label{58}
\eeq
when the field equations are satisfied.  Decomposing $\eta$ in the manner 
\beq
\eta^\alpha = \eta^\beta\nabla_\beta t\, k^\alpha + \hat\eta^\alpha , 
\label{59}
\eeq
with $\hat\eta^\alpha\nabla_\alpha t=0$, and using $\Lie_k\eta^\alpha =0$, 
we have $\delta_\eta {\cal L} = \Lie_\eta{\cal L} = \nabla_\alpha ({\cal
L}\hat\eta^\alpha )$,
\beq
\delta ( Q_L-\sum_i Q_{Li}) = \int_\Sigma \partial_\alpha ({\cal
L}\hat\eta^\alpha )d^3x = 0 , \label{60}
\eeq
because ${\cal L}$ vanishes outside the fluid (on $\partial\Sigma$).

	Similarly, from Eq.\ (\ref{35}),
\beq
\delta (Q_K-\sum_i Q_{Ki}) = -\frac{1}{8\pi} \delta \int_\Sigma
\Rab k^\beta dS_\alpha . \label{61}
\eeq
Again, for a gauge transformation that respect the Killing symmetry,
the right side is an integral over the boundary $\partial\Sigma$ of a
quantity that vanishes outside the fluid.

	Lastly, we verify the asertion made previously, that $Q$ is independent of
the 2-surface $S$ on which it is evaluated, if $S$ encloses the fluid and
any black holes.  This is immediate for $Q_K$ from Eq.\ (\ref{35}) and 
(\ref{45}).  For
$Q$ (and $ Q_L$), it follows from the fact that $Q=Q_K$ at $\lambda
=0$, together with the implication of Eq.\ (\ref{51}) that
$\frac{dQ}{d\lambda} = \delta Q$ is independent of $S$ along any sequence
of equilibria ${\cal Q}(\lambda )$.

\subsection{First law in Hamiltonian framework}
\label{sec:ham}
In applying the first law to spacetimes that are spatially conformally
flat, we will need to write it in a 3+1 form, with  metric
$\gamma_{ab}$ on $\Sigma$ and its conjugate momentum $\pi^{ab}$ as
independent variables.  Until Eq.~(\ref{eq:div}) of this section, the vector field 
$k^\alpha$ that generates time evolution is not assumed to be a Killing 
vector.\\  

Let $\Sigma = \Sigma_0$ be a Cauchy surface transverse to $k^\alpha$, 
and let $\Sigma_t = \chi_t(\Sigma)$, with $\chi_t$ the family of diffeos 
generated by $k^\alpha$. Denote by $\gamma_{ab}(t)$ the spatial metric on 
$\Sigma_t$.    Let $n^\alpha$ be 
the future-pointing unit normal to this foliation, 
and recall that one can identify spatial tensors on $\Sigma_t$ with 
spacetime tensors that are orthogonal on all of  their indices to $n^\alpha$. 
In particular,
the projection  $\gamma_{\alpha\beta}$ orthogonal to $n_\alpha$,
\beq
\label{eq:gamma}
 \gamma_{\alpha\beta} = g_{\alpha\beta} + n_\alpha n_\beta,
\eeq  
is the 4-tensor associated with the family of 3-metrics $\gamma_{ab}(t)$ 
on the slices $\Sigma_t$.   Although $k^\alpha$ is not everywhere timelike,
the fact that it is transverse to a family of spacelike hypersurfaces
means that we can introduce a nonvanishing lapse $\alpha$ and a shift 
$\omega^\alpha$ that relate $\partial_t \equiv k^\alpha$ to $n^\alpha$ 
in the usual way,  
\beq
 k^\alpha = \alpha n^\alpha + \omega^\alpha, \qquad \omega^\alpha n_\alpha = 0.
\label{eq:ka}\eeq
Then, in a chart $\{t,x^i\}$ for which $\Sigma_t$ is a $t=$constant 
surface, the metric 
$g_{\alpha\beta} =  \gamma_{\alpha\beta} - n_\alpha n_\beta$ has the form 

\beq
\label{eq:metric}
ds^2 = -\alpha^2 dt^2 +\gamma_{ij}(dx^i+\omega^i dt)(dx^j+\omega^j dt).
\eeq
With $D_a$ the covariant derivative of the spatial metric 
$\gamma_{ab}$, the extrinsic curvature of $\Sigma_t$ is given by
\beq
K_{ab}	= -\frac12\Lie_n \gamma_{ab} 
		=\frac1{2\alpha} (-\dot \gamma_{ab} 
		        +D_a \omega_b + D_b \omega_a),
\label{eq:kab}
\eeq 
where $\dot\gamma_{ab}$ is the pullback to $\Sigma$ of 
$\Lie_k\gamma_{\alpha\beta}$, vanishing when $k^\alpha$ is a Killing vector.\\

By taking as independent variables the quantities 
$\pi^{ab},\gamma_{ab},\alpha,$, with 
\beq
\pi^{ab} = - (K^{ab}-\gamma^{ab} K)\gamma^{1/2},
\label{piab}
\eeq
we now generalize the derivation of the first law to permit independent
variations of $\delta\pi^{ab},\delta\gamma_{ab},\delta\alpha,\delta\omega^a$.
 
In terms of Hamiltonian metric variables, the gravitational lagrangian 
density takes the form \cite{MTW}
\beq
 R\sqrt{-g} =  \pi^{ab}\dot\gamma_{ab}
	- \alpha{\cal H}_G - \omega_a {\cal C}_G^a 
	+ D_a (-2D^a\alpha\gamma^{\frac12} - 2\omega^b\pi^a{}_b + \omega^a\pi)
	-\dot\pi,
\label{eq:lg}
\eeq 
where 
\begin{eqnarray}
 {\cal H}_G &:=&-2 G^{\alpha\beta}n_\alpha n_\beta\, \gamma^{\frac12}  
 	     = \,-\ {}^{3}\! R \gamma^{\frac12}
 		+ (\pi_{ab}\pi^{ab} - \frac12\pi^2)\gamma^{-\frac12}, 
\label{eq:ham}\\ 
 {\cal C}_G^a &:=& -2  G^{\alpha\beta} \gamma_{\alpha}{}\!^a 
n_\beta\, \gamma^{\frac12}
  		= -2 D_b \pi^{ab}. 
\label{eq:mom}
\end{eqnarray}

Regarding $\displaystyle {\cal L}= (\frac1{16\pi}R-\epsilon)\sqrt{-g}$
as a function of $\pi^{ab},\gamma_{ab},\alpha,\omega^a$ and the fluid 
variables, we rewrite Eq. (\ref{dl1}) in the manner

\begin{eqnarray}
\delta {\cal L} &=& - \alpha\gamma^{\frac12}\rho T\Dl s 
	    - \frac{h}{u^t}\Dl (\rho u^t\, \alpha\gamma^{\frac12})
\nonumber\\ 
	    &&+ \frac1{16\pi}\{ 
	-\delta\alpha {\cal H} - \delta\omega^a {\cal C}_a 
	+ \delta \pi^{ab} [\dot\gamma_{ab}-D_a\omega_b - D_b\omega_a 
	-2\alpha(\pi_{ab} -\frac12\gamma_{ab}\pi)\gamma^{-\frac{1}{2}} ]
	\nonumber\\
&&\phantom{+ \frac1{16\pi}} -\delta\gamma_{ab}(G^{ab}-8\pi S^{ab})
\alpha\gamma^{\frac12}\}
  - \xi_\alpha\na_\beta\Tabu\alpha\gamma^{\frac12} 
  + D_a\tilde\Theta^a \gamma^{\frac12}
 - \frac1{16\pi}(\delta \pi^{ab}\, \gamma_{ab})^{\bullet} \ .
\label{eq:lham}\end{eqnarray}
Here, denoting the pullback to $\Sigma$ of $\sigma_\alpha$ by 
$\gamma_a^\alpha\sigma_\alpha$, we have set
\beq
\rho_H := T_{\alpha\beta}n^\alpha n^\beta, \ \ 
    j_a:= - T_{\alpha\beta} \gamma_a^\alpha n^\beta, \ \ 
    S_{ab}:= T_{\alpha\beta} \gamma_a^\alpha\gamma_b^\beta, 
\label{eqadmtab}
\eeq
\beq
 {\cal H} := {\cal H}_G +16\pi \rho_H\gamma^{\frac12}, \qquad 
 {\cal C}^a := {\cal C}_G^a  - 16\pi j^a\gamma^{\frac12};
\eeq
and the remaining quantities in the last line of (\ref{eq:lham}) are 
given in terms of $(\pi^{ab},\gamma_{ab}, \alpha,\omega^a)$ by

\begin{eqnarray}
  \tilde\Theta^a = \frac{1}{16\pi} \{ [&-& 2\delta (D^a\alpha\, \gamma^{1/2})
+ (\omega^a\gamma_{bc}\delta\pi^{bc} + \pi\delta\omega^a -
2\pi^a{}_b \delta\omega^b)] \gamma^{-\frac{1}{2}}\nonumber\\
&+& (\gamma^{ac}\gamma^{bd} - \gamma^{ab}\gamma^{cd})(\alpha
D_b\delta\gamma_{cd} -D_b\alpha\,\delta\gamma_{cd})\nonumber\\
&+& \alpha (\epsilon +p)q^a{}_b\xi^b - \alpha\omega^aj^b\xi_b\} , 
\label{eq:tildetheta}\end{eqnarray}

\begin{eqnarray}
G^{ab} = \dot\pi^{ab}\alpha^{-1}\gamma^{-\frac{1}{2}} &+& {}^3\!R^{ab} -
 \frac{1}{2} \gamma^{ab}{} \, ^3\!R + (2\pi^{ac} \pi^b{}_c
 -\pi\pi^{ab} - \frac{1}{2} \gamma^{ab}\pi^{cd}\pi_{cd} 
 + \frac{1}{4}\gamma^{ab}\pi^2)\gamma^{-1}  \nonumber\\
 &-& \frac{1}{\alpha} (D^aD^b\alpha -\gamma^{ab}D^2\alpha )
 +\frac{2}{\alpha} \pi^{c(b}D_c\omega^{a)}\gamma^{-\frac{1}{2}} 
 -\frac1\alpha D_c(\pi^{ab}\omega^c)\gamma^{-\frac{1}{2}},
\label{eq:Gab}
\end{eqnarray}
and
\beq
 \alpha\xi_\beta\nabla_\alpha T^{\alpha\beta} = \xi_b [D_\alpha (\alpha
T^{ab}) +D^b\alpha\, \rho_H - j_aD^b\omega^a - D_a(\omega^aj^b)] .
\eeq

For $k^\alpha$ a Killing  vector and  $A^\alpha$ any vector field Lie
derived by $k^\alpha$, we have the identities
\beq
\nabla_\alpha A^\alpha {\sqrt{-g}} = D_a\tilde A^a {\sqrt{\gamma}} ,
\label{eq:div}\eeq

\beq
\int_{\pa\Sigma} (k^\alpha A^\beta -k^\beta A^\alpha )dS_{\alpha\beta} =
\int_{\pa\Sigma} \tilde A^a dS_a ,
\eeq
where
\beq
\tilde A_a = \alpha A_\alpha \gamma^\alpha_a+\omega_aA_\alpha n^\alpha ,
\eeq
and $dS_a$ is along the outward normal to $\pa\Sigma$ in $\Sigma$.  In
particular, the vector $\tilde\Theta_a$ of Eq. (\ref{eq:tildetheta}) is related 
to $\Theta^\alpha$ by 
$\tilde\Theta_a = \alpha\Theta_\alpha \gamma^\alpha_a 
		+\omega_a \Theta_\alpha n^\alpha$, 
implying
\beq
\delta Q_L = \int_S \tilde\Theta^a dS_a, \hspace{4mm} \delta Q_{Li} =
- \int_{{\cal B}_i}\tilde\Theta^a dS_a .
\label{eq:dql}
\eeq
$Q_K$ can be expressed in terms of $(\pi^{ab}, \gamma_{ab} , \alpha ,
\omega^a)$ by writing
\begin{eqnarray}
\nabla_\alpha k_\beta\gamma^\alpha_a n^\beta &=&
\gamma_a^\alpha\nabla_\alpha (k_\beta n^\beta
)-\gamma_a^\alpha\nabla_\alpha n_\beta\, k^\beta\nonumber\\
&=& -D_a\alpha + K_{ab}\omega^b ,
\end{eqnarray}
with $K_{ab} =-(\pi_{ab}-\frac{1}{2} \gamma_{ab}\pi
)\gamma^{-\frac{1}{2}}$. Then
\begin{eqnarray}
Q_K-\sum_i Q_{Ki} &=& \frac{1}{8\pi} \int_{\pa\Sigma} (D^a\alpha
-K^a{}_b\omega^b)dS_a\nonumber\\
&=& -\frac{1}{8\pi} \int_\Sigma R^\alpha{}_\beta k^\beta dS_\alpha ,
\label{eq:qkab}
\end{eqnarray}
with\cite{det89}
\beq
R^\alpha{}_\beta k^\beta n_\alpha|_\Sigma = D_a(D^a\alpha -K^a{}_b\omega^b).
\label{eq:rkn}\eeq

We can verify directly that $R^\alpha{}_\beta k^\beta n_\alpha$ takes the
form (\ref{eq:rkn}), when written in Hamiltonian variables, using the 
Hamiltonian forms already given for ${\cal H}_G$, ${\cal
C}_{Ga}$, and $G^{ab}$.  Eq.\ (\ref{eq:ka}) implies
\beq
\left. R^\alpha{}_\beta k^\beta n_\alpha\right|_\Sigma = -\frac{1}{4}
\alpha\gamma^{-\frac{1}{2}}{\cal H}_G -\frac{1}{2}\gamma^{-\frac{1}{2}} 
{\cal C}_{Ga}\omega^a + \frac{1}{2} \alpha\gamma_{ab} G^{ab}. 
\label{eq:rnk}
\eeq
Eq.\ (\ref{eq:Gab}) gives
\beq
\gamma_{ab} G^{ab}= - \frac{1}{2} {} ^{(3)}R + \frac{2}{\alpha} D^2\alpha 
 + \gamma^{-1} (\frac{1}{2} \pi^{ab}\pi_{ab} - \frac{1}{4} \pi^2)
 + \frac{2}{\alpha} \pi^{ab}D_a\omega_b\gamma^{-\frac{1}{2}}
-\frac{1}{\alpha} D_a(\pi\omega^a )\gamma^{-\frac{1}{2}} ;
\eeq

and substituting this and the forms  (\ref{eq:ham}) and (\ref{eq:mom}), 
of ${\cal H}_G$ and ${\cal C}_{Ga}$ in Eq.~(\ref{eq:rnk}), we obtain
\beq
\left.R^\alpha{}_\beta k^\beta n_\alpha \right|_\Sigma = D_a(D^a\alpha
 + \pi^a{}_b\omega^b\gamma^{-\frac{1}{2}} 
 -\frac{1}{2} \pi\omega^a \gamma^{-\frac{1}{2}} ) 
 = D_a(D^a\alpha -K^a{}_b\omega^b).
\eeq

Consequently,
Eq.\ (\ref{dqk}) holds with $R$ and $G^\alpha{}_\beta n_\alpha$ given 
by Eqs.~(\ref{eq:lg}), (\ref{eq:ham}), and (\ref{eq:mom}), and with 
$\Delta g_{\alpha\beta}$ defined as a function of
$(\gamma_{ab}, \alpha , \omega^a)$, independent of $\pi^{ab}$.

Finally, combining Eq.\ (\ref{dqk}), Eq.\ (\ref{eq:lham}) and 
Eq.\ (\ref{eq:dql}) in the lagrangian derivation, as we obtain 
Eq.\ (\ref{eq:dQ0}), 
\begin{eqnarray}
\delta \left(Q - \sum_i Q_i\right) &=& \int_\Sigma ({\bar{T}} \Delta dS
+ {\bar{\mu}} \Delta dM_B + v^a\Delta dC_a)
\nonumber\\
&& \mbox{} +  \frac{1}{16\pi} \int_\Sigma\ \{ \delta\pi^{ab}
[D_a\omega_b+D_b\omega_a +2\alpha (\pi_{ab} -
\frac{1}{2} \gamma_{ab}\pi )\gamma^{-\frac{1}{2}}]
- \alpha \delta{\cal H} - \omega^a \delta{\cal C}_a \nonumber\\
&& \hspace{1cm} 
\ + \ \alpha[\delta\gamma_{ab}(G^{ab}-8\pi S^{ab}) 
+ 16\pi\, \xi_\alpha\nabla_\beta T^{\alpha\beta}]\gamma^{\frac{1}{2}}\,\}d^3x. 
\label{hamlaw}
\end{eqnarray}
Here the last two integrals in Eq.\ (\ref{eq:dQ0}) are combined by using 
$ 2\delta[(G^\alpha{}\!_\beta - 8\pi T^\alpha{}\!_\beta)k^\beta
n_\alpha] = \delta \alpha {\cal H} + \alpha \delta{\cal H} 
+ \delta \omega^a {\cal C}_a + \omega^a \delta {\cal C}_a $.

When the field equations are satisfied, and $\pi^{ab}$ is given by
\beq
\pi^{ab}=- \frac1{\alpha} (D^{(a}\omega^{b)} -
\gamma^{ab}D_c\omega^c)\gamma^{1/2},
\label{eq:pi}\eeq
we have
\beq
\delta Q = \int_\Sigma 
\left[{\bar{T}} \Delta dS + {\bar{\mu}} \Delta dM_B
+ v^a\Delta d C_a \right] + \sum_i \kappa_i\delta A_i
\eeq
 
\section{Application to the inspiraling binary black hole -- neutron star 
system}
\label{sec:4}

\subsection{Comparing configurations in quasi-stationary systems}

   Our study of a generalized first law was spurred by the fact that
equilibria stationary in a rotating frame -- spacetimes with helical
Killing vectors -- are used in several approaches to binary
inspiral.  In each of these cases, one
approximates the inspiral phase of binary coalescence by an
evolutionary path through a sequence of equilibria.  The first law has
a strikingly simple form when used to compare such dynamically related
spacetimes:  For isentropic fluids, dynamical evolution conserves the
baryon mass, entropy, and vorticity of each fluid element, and we show
that the first law becomes
\beq
   \delta Q = \frac1{8\pi}\sum_i \kappa_i\delta A_i;
\label{eq:dQ2}\eeq
or 
\beq
 \delta Q = 0,
\eeq
for perfect fluid spacetime with no black holes.
In the gauge that we have chosen ($\delta k^\alpha = 0$), when the 
spacetime is asymptotically flat and $k^\alpha$ has the asymptotic 
form $t^\alpha + \Omega \phi^\alpha$, with $t^\alpha$ and $\phi^\alpha$ 
timelike and rotational Killing vectors of a flat asymptotic metric, 
we find 
\beq
	 \delta Q = \delta M - \Omega\delta J,
\eeq
with $M$ and $J$ the mass and angular momentum at spatial infinity.  
In particular, the first law in this form describes \\
(i) comoving binaries, flows with $v^\alpha = 0$; and
(ii) irrotational binaries, potential flows $hu_\alpha = \nabla_\alpha \Phi$,
with 
\beq
\Delta (hu_\alpha) = \nabla_\alpha \Delta\Phi.
\label{dcirc}
\eeq  
   
  For an isentropic fluid, conservation of rest mass, entropy, and vorticity 
have the form 
\beq
	\Lie_u(\rho\sqrt{-g})= 0,\ \ \ 
	\Lie_u s = 0, \ \ \ 
	\Lie_u \omega_{\alpha\beta}=0,
\eeq
with the relativistic vorticity $\omega_{\alpha\beta}$ given by
\beq
\omega_{\alpha\beta} = 
 q_\alpha{}^\gamma q_\beta{}^\delta\left[\na_\gamma \left(h u_\delta\right)  
          - \na_\delta \left(h u_\gamma\right)\right] 
	= \na_\alpha \left(h u_\beta\right) - 
        \na_\beta \left(h u_\alpha\right).           
\eeq

The perturbed conservation laws have the first integrals
\beq
\Dl (\rho u^\alpha \sqrt{-g})=0, \ \ \ 
\Dl s = 0 \ \ \  \Dl\omega_{\alpha\beta}=0,
\eeq
appropriate to the difference between two flows that are related 
by a dynamical evolution.  
It immediately follows that the first and second terms of Eq.(\ref{eq:dQ}) 
vanish for isentropic flows. \\

	To see that the third term vanishes 
when the perturbed vorticity vanishes, we use $[d,\Lie_\xi] = 0$ to write
\beq
 0 = \Delta \omega_{\alpha\beta} = 
  \na_\alpha \Delta\left(h u_\beta\right) - 
        \na_\beta\Delta \left(h u_\alpha\right),        
\eeq
implying $\Delta{hu_\alpha} =\nabla_\alpha \Delta\Phi$, as in 
Eq. (\ref{dcirc}).  
 The third term in Eq. (\ref{eq:dQ}) can then 
be written
\beqn
 \int_\Sigma v^\beta\Dl(hu_\beta)\rho u^\alpha\dSa 
 &=& \int_\Sigma v^\beta\na_\beta\Delta\Phi \rho u^t k^\alpha\dSa \\ 
 &=& \int_\Sigma \na_\beta (v^\beta\Delta\Phi \rho u^t) k^\alpha\dSa
     - \int_\Sigma \na_\beta (v^\beta \rho u^t)\Delta\Phi k^\alpha\dSa
\eeqn
The first term in this last equality vanishes, because it is the integral
of a 
total divergence. (Write   
$(\na_\beta A^\beta)k^\alpha = \na_\beta(A^\beta k^\alpha - A^\alpha
k^\beta)$ 
and use	Stokes' theorem; or, more concretely, write 
 $k^\alpha dS_\alpha = \sqrt{-g} d^3x$.)  For the second term, recalling the 
definition of $v^\alpha$ in Eq. (\ref{eq:udec}), we have
\beq
\na_\beta (v^\beta \rho u^t) = \na_\beta (\rho u^\beta) -
\Lie_k(\rho u^t)
 - \rho u^t\na_\beta k^\beta,
\eeq		
with each term on the right separately vanishing.

Thus, for spacetimes related by a perturbation that locally conserves baryon 
mass, entropy and vorticity, the first law has the form 
(\ref{eq:dQ2}), as claimed.

\subsection{Asymptotically flat systems}
\label{sec:af}

We will use the 3+1 formalism of Sec.~(\ref{sec:ham}) to evaluate 
$\delta Q = \delta Q_K + \delta  Q_L$.  In the post-Newtonian and in the 
Isenberg-Wilson-Mathews spacetimes that have been used to describe 
binary systems, the 3-metric has the asymptotic form
\beq 
	\gamma_{ab} = f_{ab} + O(r^{-1}),
\label{eq:a1}
\eeq
where $r = (\delta_{ij}x^i x^j )^{1/2}$, with $\{ x^i\}$ a chart 
for which $f_{ij} = \delta_{ij}$.  \\ 

By writing  $k^\alpha = \alpha n^\alpha + \omega^\alpha$, as in 
Sec.~(\ref{sec:ham}), 
we choose a shift $\omega^a$
associated with a comoving chart at spatial infinity.  That is, 
\beq 
 \omega^a = \Omega \phi^a + \beta^a,\ \ \ \mbox{where}\ \ \ \beta^a = O(r^{-2}), 
\label{eq:a2}
\eeq
and $\phi^a$ is a rotational Killing vector of the flat metric 
$f_{ab}$
\beq 
\phi^a = x^1(\partial_2)^a - x^2 (\partial_1)^a .
\label{eq:phi}
\eeq
The extrinsic curvature and lapse have asymptotic behavior
\beq
K_{ab} = O(r^{-3}),\qquad
\alpha = 1 + O(r^{-1}), \qquad  D_a\alpha = O(r^{-2}). \\
\label{eq:a3}
\eeq

To evaluate $\delta Q$, we first define 
two asymptotic masses and the asymptotic angular momentum.   
A mass $M_K$ seen by a test particle in Keplerian orbit is associated 
with the asymptotic form of the lapse,   
\beq
M_K :=  \frac1{4\pi}\int_\infty D^a\alpha\ dS_a
 = \lim_{r\rightarrow\infty} \frac1{4\pi}\int_{S_r} 
\partial_r\alpha\  r^2 d\Omega,
\label{eq:mk}
\eeq
where 
\[ \int_\infty :=\lim_{r\rightarrow\infty}\int_{S_r}  \]
with $S_r$ a sphere of constant $r$.  
In terms of the metric potentials, $M_K$ has the form of the  
Komar mass associated with an timelike asymptotic Killing vector 
$t^\alpha$.\\

The ADM mass is computed from the 3-metric: 
\begin{eqnarray}
\madm &=&\ \frac1{16\pi}\int_\infty
 (f^{ac}f^{bd}-f^{ab}f^{cd})\partial_b\gamma_{cd}\  dS_a
\nonumber\\ 
  &=& -\lim_{r\rightarrow\infty} \frac1{2\pi}\int_{S_r} 
\partial_r \psi\  r^2 d\Omega.
\label{eq:madm}
\end{eqnarray}	
Finally, the angular momentum associated with the asymptotic rotational 
Killing vector  is 
given by 

\beq
 J = -\frac1{8\pi}\int_\infty \pi^a{}_b\ \phi^b dS_a 
   = \frac1{8\pi}\int_\infty K^a{}_b\ \phi^b dS_a .
\label{eq:j}
\eeq

As in the first equality of Eq.~(\ref{eq:qkab}), we have
\beq
Q_K = -\frac1{8\pi}\int_\infty ( - D^a\alpha + K^a{}_b \omega^b)dS_a
	= \frac1{8\pi}\int_\infty ( D^a\alpha - K^a{}_b \Omega\phi^b)dS_a,
\eeq
whence, by Eqs. (\ref{eq:mk}) and (\ref{eq:j})
\beq
	Q_K = \frac12 M_K -\Omega J.
\label{eq:qk}
\eeq

We obtain $\delta Q_L$ from Eqs. (\ref{eq:dql}) and (\ref{eq:tildetheta}). 
Using $\phi^a D_a r =0$ and the asymptotic behavior given above, we have
\begin{eqnarray}
  \tilde\Theta^a = \frac{1}{16\pi} \{[&-& 2 f^{1/2} \delta D^a \alpha\, 
	-2 \delta(\pi^a{}_b\,\Omega\phi^b ) +2\delta\pi^a{}_b\, \Omega\phi^b
]f^{-1/2}
\nonumber\\
	&+& (f^{ac} f^{bd} - f^{ab} f^{cd})D_b\delta\gamma_{cd} \} 
	 + O(r^{-3}), 
\end{eqnarray}

\begin{eqnarray}
  \delta\tilde Q_L = \frac{1}{16\pi} [&-& 2\delta \int_\infty D^a \alpha\ dS_a
	+2 \delta\int_\infty K^a{}_b\Omega\phi^b dS_a 
	- 2\Omega\ \delta \int_\infty K^a{}_b \phi^b dS_a
\nonumber\\
	&+& \int_\infty (f^{ac} f^{bd} - f^{ab} f^{cd})D_b\delta\gamma_{cd} dS_a\,]
\nonumber\\
 = && -\frac12 \delta M_K + \delta(\Omega J) - \Omega\delta J + \delta \madm
\nonumber\\
 =&&  \delta M_{\rm ADM} -\frac12 \delta M_K + \delta \Omega  J. 
\label{eq:dql2}
\end{eqnarray}   
Adding Eq. (\ref{eq:dql2}) to $\delta$(\ref{eq:qk}), we have 
\beqn
\delta Q &=& \frac12 \delta M_K -\delta(\Omega J)
	    + \delta M_{\rm ADM} -\frac12 \delta M_K + \delta \Omega  J\nonumber\\
	&=& \delta M_{\rm ADM} - \Omega \delta J,
\eeqn
in agreement with the usual first law.

\subsection{The first law for spacetimes with a conformally flat 
	    spatial geometry}
\label{sec:scf}
As mentioned earlier, several groups have recently obtained
quasi-equilibrium sequences\cite{ba97,bgm989,mmw99,use00},
approximating binary inspiral by a sequence of Isenberg-Wilson-Mathews
spacetimes (henceforth {\em IWM spacetimes}), spacetimes whose
3-geometry is conformally flat and whose five metric potentials satisfy
a truncated set of five field equations.  More precisely, the metric of
a IWM spacetime satisfies the constraint equations
and the spatial trace of the Einstein equation, together with the
maximal slicing condition for its conformally flat slices; and its
matter satisfies the equation of motion, $\na_\beta T^{\alpha\beta} =
0$ (see e.g., Isenberg\cite{isenberg} or Flanagan\cite{flanagan}).
\\

As Detweiler has pointed out, when the spacetime has a helical  (or
timelike) Killing vector, one cannot in general solve all of these
equations simultaneously for a metric with conformally flat spacelike
slices. One must omit one  relation to accommodate the new constraint
that the existence of a Killing  vector imposes on the extrinsic
curvature, $K_{ab}$. We note first that, if one omits the $K=0$ condition,
the resulting spatially conformally flat spacetime satisfies an exact first
law, despite the fact that only a truncated set of field equations are
imposed.\\

In the second part of this section, we note that one can alternatively
retain the $K=0$ condition if one simply {\it defines} a tensor $\hat K_{ab}$
by the form (Eq.( \ref{eq:cfkij}) below) that the  extrinsic 
curvature would take in a spacetime with a helical Killing vector 
foliated by $K=0$ slices. We show that the first 
law is {\it exact} in this framework.  This is surprising, in view of 
the artificiality of the definition of  $\widehat K_{ab}$ and the fact
that one component of the Einstein equation is not satisfied in
the IWM framework.\\

In each case, one has a spacetime foliated by hypersurfaces whose spatial
metrics have the form
\beq
\gamma_{ab} = \psi^4f_{ab},
\eeq
with $f_{ab}$ a flat metric.  The corresponding 4-tensors, 
\beq
\gamma_{\alpha\beta} = \psi^4f_{\alpha\beta},
\eeq
are Lie derived by the Killing vector $k^\alpha$:
\beq
\Lie_k \gamma_{\alpha\beta} = 0, \hspace{3mm} \Lie_k \psi =0,
\hspace{3mm} \Lie_k f_{\alpha\beta} = 0.
\eeq
In particular (although we will not use the fact in this section),
\beq
k^{\alpha}=t^{\alpha}+\Omega\phi^{\alpha},
\label{eq:ka1}
\eeq
with $\phi^a$ a rotational Killing vector of $f_{ab}$.

In the first case (with $K$ not required to vanish), the spacetime
satisfies on each $\Sigma_t$ the equations
\beq
{\cal H} = 0, \hspace{2mm} {\cal C}_a =0, \hspace{2mm}
\gamma_{ab}(G^{ab}-8\pi\, T^{ab})=0, \hspace{2mm} \nabla_\beta
T^{\alpha\beta} =0, \label{scf}
\eeq
together with the relation (\ref{eq:pi}) expressing $\pi^{ab}$ in terms of the
metric.  Because
\beq
\delta\gamma_{ab} = 4 \frac{\delta\psi}{\psi} \gamma_{ab},
\eeq
it is exactly this set of equations that occur in the action and in the 3+1
form of the first law (\ref{hamlaw}), when one compares two spatially
conformally flat spacetimes.

Finally, comparing asymptotically flat spacetimes of this kind, with no
local change in entropy, baryon number, or circulation, we have
\beq
\delta M = \Omega\delta J + \sum \kappa_i\delta A_i . \label{dm}
\eeq

We consider next solutions $(\hat\pi^{ab}, \gamma_{ab}, \alpha , \omega^a ,
\epsilon , u^\alpha )$, to the same set (\ref{scf}) of equations, now with
$\hat\pi =0$:
\beq
\hat\pi^{ab} =-\hat K^{ab} \gamma^{\frac12} ,
\eeq
with $\hat K_{ab}$ the {\it tracefree part} of the extrinsic curvature:
\beq
\hat K_{ab} = \frac{1}{2\alpha} (D_a\omega_b + D_b\omega_a - \frac{2}{3}
\gamma_{ab}D_c\omega^c). \label{eq:cfkij}
\eeq

	One writes ${\cal H}$, ${\cal C}_a$, 
$\gamma_{ab}(G^{ab}-8\pi\, T^{ab})$,
and $\nabla_\beta T^{\alpha\beta}$ as they occur in the Hamiltonian
formalism (for the metric), as functions of $(\pi^{ab}, \gamma_{ab}, \alpha
, \omega^a)$ and the matter variables; one substitutes for $\gamma_{ab}$
and $\pi^{ab}$ the expressions
\beq
\gamma_{ab} = \psi^4f_{ab}, \hspace{4mm}
\hat\pi^{ab}=-\frac{1}{2\alpha} (D^a\omega^b +
D^b\omega^a - \frac{2}{3} \gamma^{ab} D_c\omega^c)\gamma^{-\frac12};
\eeq
and one solves the resulting system of equations for $(\psi , \alpha ,
\omega^a; \epsilon , v^a)$.
\beq
\hat{\cal H}=0, \hspace{3mm} \hat{\cal C}_a=0, \hspace{3mm}
	( \hat G^{ab}-8\pi S^{ab})\gamma_{ab}=0, \hspace{3mm}
 	\nabla_\beta T^{\alpha\beta} =0,
\eeq
where
\beq
\hat{\cal H} = {\cal H}(\hat\pi^{ab},\gamma_{ab}, \alpha, \omega^a; 
\epsilon, u^\alpha), 
\hspace{2mm} 
\hat{\cal C}^a = {\cal C}^a(\hat\pi^{ab},\gamma_{ab}, \alpha, \omega^a; 
\epsilon, u^\alpha), 
\hspace{2mm} 
\hat G^{ab}\gamma_{ab} =
G^{ab}\gamma_{ab}(\hat\pi^{ab},\gamma_{ab}, \alpha, \omega^a; 
\epsilon, u^\alpha).
\eeq
Then, for a family of such solutions, the quantities $\delta\alpha$,
$\delta\omega$, and $\delta\gamma_{ab} = 4 \frac{\delta\psi}{\psi} \gamma_{ab}$
occurring on the right of the first law (\ref{hamlaw}) multiply expressions
that vanish.  Because $\delta\hat\pi^{ab}$ is traceless, the expression
involving $\delta\hat\pi^{ab}$ has the form
\beq
\frac{1}{16\pi} \delta\hat\pi^{ab}
[D_a\omega_b+D_b\omega_a+2\alpha\gamma^{-\frac{1}{2}}\hat\pi_{ab}] =
\frac{1}{16\pi} \delta\hat\pi^{ab} [D_a\omega_b+D_b\omega_a - \frac{2}{3}
\gamma_{ab}D_c\omega^c+2\alpha\gamma^{-\frac{1}{2}} \hat\pi_{ab}]
= 0 .
\eeq
Eq.\ (\ref{hamlaw}) thus yields
\beq
\delta\hat Q_L - \frac{1}{8\pi} \delta\int \hat R^\alpha{}_\beta k^\beta
dS_\alpha = \int_\Sigma [\overline T\Delta dS+\overline\mu\Delta
dM_B+v^a\Delta dC_a].
\eeq

   To recover the first law in the form
\beq
\delta M = \Omega\delta J ,
\eeq
we must show that
\beq
-\frac{1}{8\pi} \delta\int_\Sigma \hat R^\alpha{}_\beta k^\beta dS_\alpha =
-\frac{1}{8\pi} \delta\int_{\pa\Sigma} \nabla^\alpha k^\beta dS_{\alpha\beta} .
\eeq
This is not obvious, because, in replacing the extrinsic curvature by its
tracefree part, we invalidate the Killing identity (\ref{eq:kill}):
\beq
\nabla_\beta\nabla^\alpha k^\beta \not= R^\alpha{}_\beta (\hat\pi^{ab},
\psi , \alpha , \omega^a ) k^\beta
\eeq
Remarkably, however, the $n_\alpha$ components of the two sides of this
inequality differ by a divergence; and the asymptotic behavior of the
spacetime implies the equality
\beq
Q_K =-\frac{1}{8\pi} \int_\infty \nabla^\alpha k^\beta dS_{\alpha\beta} =
     -\frac{1}{8\pi} \int \hat R^\alpha{}_\beta k^\beta dS_\alpha
\label{eq:qkcf}
\eeq
That is, from Eq.\ (\ref{eq:rkn}), we have
\beq
\left.\nabla_\beta\nabla^\alpha k^\beta n_\alpha \right|_\Sigma =
R^\alpha{}_\beta k^\beta n_\alpha |_\Sigma = D_a(D^a\alpha
-K^a{}_b\omega^b);
\eeq
and
\beq
\left.\hat R^\alpha{}_\beta k^\beta n_\alpha \right|_\Sigma =
D_a(D^a \alpha -\hat K^a{}_b\omega^b) .
\eeq
Then
\begin{eqnarray}
\left.\nabla_\beta\nabla^\alpha k^\beta n_\alpha \right|_\Sigma - \hat
R^\alpha{}_\beta k^\beta n_\alpha |_\Sigma &=& D_a[(\hat K^a{}_b
-K^a{}_b)\omega^b]\nonumber\\
&=& - \frac{1}{3} D_a(\omega^aK).
\end{eqnarray}
As noted in Sect.~(\ref{sec:af}),
\beq
\omega^a = \Omega \phi^a + \beta^a,\ \mbox{ with }\ \beta^a = O(r^{-2}),
\eeq
where $\phi^a$ is a rotational Killing vector of the flat metric $f_{ab}$;
and $K = O(r^{-3})$.
\footnote{ If, however, one allows a nonzero 3-momentum, with
boosted-Schwarzschild asymptotics, then $\psi = 1+ f(\hat r)/r +
O(r^{-2})$ and and $\beta^a = O(r^{-1})$.  Because $K$ is given by 
$\displaystyle{\frac1{\alpha}(\frac6{\psi}\Omega
\phi^a D_a\psi + D_a\beta^a)}$, one can only
demand $K=O(r^{-1})$, $\beta^a K = O(r^{-2})$, allowing a finite
contribution to $Q_K$. } We then have
\begin{eqnarray}
Q_K &=& -\frac{1}{8\pi} \int R^\alpha{}_\beta k^\beta dS_\alpha =
-\frac{1}{8\pi} \int_\Sigma \hat R^\alpha{}_\beta k^\beta dS_\alpha -
\frac{1}{24\pi} \int_{\pa\Sigma} \omega^b K dS_b\nonumber\\
&=& - \frac{1}{8\pi} \int_\Sigma \hat R^\alpha{}_\beta k^\beta dS_\alpha ,
\end{eqnarray}
as claimed. \\

	From Eq.~(\ref{eq:madm}), we conclude
\beq
\delta M =\Omega\delta J,
\eeq
along a family of conformally flat solutions to the IWM equations, written
in terms of $(\hat\pi^{ab}, \psi , \alpha , \omega^a)$.\\

   Note that the equation $\xi_\beta\na_\alpha\Tabu  = 0$ is satisfied,
because, for an isentropic fluid, the equation of hydrostatic
equilibrium, conservation of rest-mass, and the one-parameter equation
of state together imply $\na_\alpha\Tabu  = 0$. To see this explicitly,
we decompose the divergence of the stress tensor as follows:

\beq
	\na_\alpha\Tabu
	= (q_{\beta\gamma} - u_\beta u_\gamma)
	\na_\alpha T^{\alpha\gamma}
	= \rho
	\left[u^\gamma\na_\gamma(hu_\beta) +\na_\beta h \right]
	+ u_\beta h \na_\alpha(\rho u^\alpha)
	- \rho T  \na_\beta\, s
\label{eq:EOM}
\eeq
In constructing an isentropic ($s={\rm const}$) equilibrium model,
conservation of rest mass is assumed, and a barotropic equation of
state $p=p(\rho)$ is used.  Helical symmetry and the assumption that
the fluid flow is either co-rotational or irrotational then leads to a
first integral of the Euler equation $u^\gamma\na_\gamma(hu_\beta)
+\na_\beta h=0$.  It is this first integral, specialized to a
conformally flat metric, that is solved in the IWM formalism, implying
that $\na_\alpha\Tabu  = 0$.  Thus, as claimed, all terms involving the
field equations vanish in Eq. (\ref{eq:dQ0}), and the first law holds
for IWM spacetimes in the form (\ref{eq:dQ}). \\

As in the exact theory, when the system includes black holes, the
$\kappa_i \delta A_i$ terms refer to Killing horizons.  The IWM
spacetimes do not satisfy the Raychaudhuri equation for the null
generators of the horizon; as a result, as noted in the introduction,
Killing horizons in IWM spacetimes need not co-rotate with the orbital
motion.  

\section{Discussion}  

	The first law can be used to deduce a criterion for orbital
stability for the asymptotically flat models of binary equilibria
discussed above, using a theorem of Sorkin\cite{turning1}.  Consider a
one-parameter family ${\cal Q}(\lambda)$ of binary equilibrium models
along which baryon number, entropy and circulation are locally constant
(the lagrangian changes $\Delta s, \Delta dM_B$, and $ \Delta
dC_\alpha$ vanish).  Suppose that $\dot J =0$ at a point $\lambda_0$
along the sequence, and that $\ddot\Omega \dot J \neq 0$ at
$\lambda_0$.  Then the part of the sequence for which $\dot\Omega \dot
J >0$ is unstable for $\lambda$ near $\lambda_0$.  \\

The result relies on a first law in the form
\beq
dM = \Omega dJ
\eeq
and on the fact that the equilibria are extrema of 
mass with $J$ constant.  As we have seen, this is the case for a configuration 
space in which baryon number, entropy, and circulation are fixed for
each fluid element. For asymptotically flat models with one or more black holes, if one also fixes 
the area of the horizon along a sequence, then the same criterion 
above can be used to diagnose stability. \\ 

In general, the proof of the theorem shows only that the spacetime is
{\em secularly} unstable on one side of the turning point.  In the present
context, however, the theorem shows the existence of nearby configurations
with lower mass that can be reached by perturbations that conserve
baryon number, entropy, and circulation; this suggests that the
criteria locates the onset of dynamical instability.\\

When one models stationary binary systems in full GR, the lack of asymptotic
flatness leads to several ambiguities.  For binary charges in Minkowski 
space, one can obtain a one-parameter family of equilibria if one simply
replaces asymptotic regularity (finite energy) by a condition that the 
electromagnetic field be given by the half-advanced + half-retarded Green's 
function.  In GR, it remains to be seen whether one can find an analogous 
asymptotic condition.  Simply requiring equal amounts of ingoing and 
outgoing radiation is a weaker condition even in Minkowski space; in 
GR one must have asymptotic conditions as restrictive as asymptotic 
flatness to avoid ambiguity in each asymptotic multipole.  Finally,
as mentioned in Sect.~\ref{sec:3}, the helical Killing vector has an 
arbitrary scaling that one must resolve to obtain a unique value for 
the charge $Q$ of the first law.

\acknowledgments
We thank Ric Ancel, Abhay Ashtekar, Steven Detweiler, 
Eric Gourgoulhon, James Isenberg, Stephen
Hawking and Robert Wald for helpful conversations; 
and we also thank Detweiler and Gourgoulhon for reading and pointing 
out errors in an earlier draft. 

This research was supported in part by NSF Grant PHY00-71044
and by a Japanese Monbu-kagakusho Grant (No. 13740143).    

%%%%%%%%%%%%%%%%%%%%%%%%%%%%%%%%%%%%%%%%%%%%%%%%%%%%%%%%%%%%%%%%%%%%%%%%%%%%%

\appendix

\newcommand{\vp}{\phi}
\def\Madm{M_{\rm ADM}}
\def\Jadm{J}
\def\Mchi{M_{\rm \chi}}
\def\2pi{2\pi}

\section{Virial relation in IWM spacetimes} 
\label{appendix1} 

In this appendix, we derive a virial relation 
for quasiequilibrium states in IWM spacetimes. 
Incidentally, we show that the virial relation is equivalent to 
the relation $M_K=\Madm$. \footnote{The relation $M_K=\Madm$ 
for stationary and asymptotically flat spacetimes has been proven 
by Beig \cite{beig}, 
and by Ashtekar and Magnon-Ashtekar \cite{aa79}. 
A virial relation relying on this has been derived by 
Gourgoulhon and Bonazzola \cite{gb94}.}  
 
As described in ~\ref{sec:scf}, we use a 3+1 formalism, with 
3-metric $\gamma_{ab} = \psi^4 f_{ab}$, and with a helical Killing 
vector that has the form
\beq
k^{\alpha}=t^{\alpha}+\Omega\phi^{\alpha},
\eeq
where $\phi^a$ is a rotational Killing vector of the flat metric $f_{ab}$. 
Throughout this appendix, 
we use Cartesian coordinates $t,\{x^i\}$ for which  $f_{ab}$ has components 
$f_{ij} = \delta_{ij}$ and $t^\mu = \delta_0^\mu$.
In the IWM formalism, one imposes the maximal slicing condition 
$K=0$ on the family of $t=$const surfaces $\Sigma_t$; and, instead of 
solving the full Einstein equation, one solves the
Hamiltonian constraint, the momentum constraint and the equation for the
slicing condition.
Here, however, as in Sect.~\ref{sec:scf}, to obtain a set of equations
consistent with the existence of a helical Killing vector, we replace
the extrinsic curvature in this set of equations by its tracefree part,
$\hat K^{ij}$.
 
The basic equations are then
\beqn
&& \Delta \psi= -2\pi \psi^5 \rho_H  - {\psi^5 \over 8} 
\hat K_{i}^{~j} \hat K_{j}^{~i}\equiv -S_{\psi},\label{eqpsi} \\
&&\pa_j (\sqrt{\gamma} \hat K_i^{~j})=8\pi j_i \sqrt{\gamma}, \label{momcon} \\
&& \Delta \chi= 2\pi \chi \psi^4 (\rho_H + 2S_k^{~k})+{7 \over 8}\chi \psi^4 
\hat K_{i}^{~j} \hat K_{j}^{~i} \equiv S_{\chi},\label{eqchi} 
\eeqn
where $\Delta$ denotes the flat Laplacian for three space, 
$\sqrt{\gamma}={\rm det}(\gamma_{ij})=\psi^6$, and 
$\chi \equiv \alpha\psi$. (See Eqs. (\ref{eqadmtab}) for 
definition of $\rho_H$, $j_i$ and $S_{ij}$.)
The energy-momentum tensor is
assumed to be nonzero only inside the light cylinder 
$(x^2+y^2)^{1/2}< \Omega^{-1}$.

The shift vector $\beta^a$ of Eq.~(\ref{eq:a2}) satisfies
\beq
\beta^\alpha = -\alpha n^\alpha + t^\alpha.
\eeq  
The Cartesian components $\hat K_i^j$ are given in terms of $\beta^i$ by
\beq
\hat K_{i}^{~j} = {1 \over 2\alpha}\biggl( \pa_i \beta^j + 
\delta_{il} \delta^{jk}
\pa_k \beta^l-{2 \over 3} \delta^j_i \pa_k \beta^k \biggr). \label{eqKij}
\eeq

The asymptotic behavior of geometric variables is that of Eqs.~(\ref{eq:a1})-
(\ref{eq:a3}),
\beqn
&&\psi = 1 + O(r^{-1}),\\
&&\chi =  1 + O(r^{-1}),\\
&&\beta^i = O(r^{-2}),\\
&&\hat K_i^{~j} = O(r^{-3}),
\eeqn
appropriate for an asymptotically flat spacetime in a 
chart for which the total ADM 3-momentum vanishes:  
\beqn
P_i \equiv 
{1 \over 8\pi} \oint_{r\rightarrow \infty} K_{i}^{~j}\sqrt{\gamma} \ds_j
=0,
\eeqn
where $\ds_j=(D_j r) r^2 d\Omega$ and $\sqrt{\gamma}$ is computed in 
Cartesian coordinates.

We now derive the virial relation and show the equivalence of 
Komar and ADM mass for quasiequilibria of two neutron stars.  
For $r\rightarrow \infty$, $\chi$ and 
$\psi$ behave as $1-\Mchi/2r + O(r^{-2})$ and 
$1 + \Madm/2r + O(r^{-2})$. 
{}From this asymptotic behavior, 
we can define $\Mchi$ and $\Madm$ by the surface integrals, 
\beqn
\Mchi &=& {1 \over \2pi} \oint_{r \rightarrow \infty} 
\delta^{ij} \psi \pa_i \chi \ds_j \nonumber \\
\Madm &=& -{1 \over \2pi} \oint_{r \rightarrow \infty} 
\delta^{ij} \chi \pa_i \psi \ds_j. 
\eeqn
Since $\Mchi=-\Madm + 2M_K$, our goal is to show $\Mchi=\Madm$.  

Using Gauss's law, they can be rewritten in the manner 
\beqn
\Mchi &=& 
{1 \over 2\pi} \int (\psi S_{\chi} + \delta^{ij} \pa_i \chi \pa_j \psi) 
d^3x, \label{eqmchi} \\
\Madm 
&=& {1 \over \2pi} \int (\chi S_{\psi} - 
\delta^{ij} \pa_i \psi \pa_j \chi) d^3x.
\label{eqmadm}
\eeqn

We next derive a relation that will be used several times in the 
calculations that follow.  From \\ 
$\chi \psi^5 \hat K_i^{~j} \hat K_j^{~i} = \sqrt{\gamma} \hat K_i^{~j} \pa_j \beta^i$, 
we have 
\beqn
& &\int \chi \psi^5 \hat K_i^{~j} \hat K_j^{~i} d^3x 
= \int \sqrt{\gamma} \hat K_i^{~j} \pa_j \beta^i d^3x \nonumber \\
&=& - \int \pa_j (\sqrt{\gamma} \hat K_i^{~j}) \beta^i d^3x 
+\oint \sqrt{\gamma} \hat K_i^{~j} \beta^i \ds_j \nonumber \\
&=& -8\pi \int \sqrt{\gamma} j_i \beta^i d^3x,
\label{eqid1}
\eeqn
where we use the asymptotic behaviors at $r \rightarrow \infty$ 
and Eq. (\ref{momcon}) to obtain the last line. 
$\oint$ without specification of a surface denotes a surface integral
over $\partial\Sigma$:
$\oint=\oint_{r\rightarrow \infty}$. 
{}From the vanishing of the total ADM 3-momentum (more precisely, 
from the vanishing of $\int_{r\rightarrow\infty}\hat K_i^j 
\sqrt{\gamma}dS_j$) and 
from the momentum constraint (\ref{momcon}), we have 
\beqn
0= \int j_i \sqrt{\gamma} d^3x, \label{linear}
\eeqn
which may be interpreted as the 
linear momentum of a neutron star.

Using Eqs. (\ref{eqmchi}) and (\ref{eqmadm}), we write the difference 
between $\Madm$ and $\Mchi$ in the form 
\beqn
\Mchi - \Madm =&&{1 \over \pi} \int \biggl[ 2\pi \chi \psi^5 S_k^{~k} 
+ {3 \over 8} \chi \psi^5 K_{i}^{~j} K_{j}^{~i} 
+\delta^{ij}\pa_i\psi \pa_j \chi  \biggr]d^3x  \nonumber \\
 = &&2 \int \biggl[ \sqrt{\gamma} 
\{j_k v^k + 3\alpha P\} - {1 \over 2}\sqrt{\gamma} j_j \beta^j 
+{1 \over 2 \pi}
\delta^{ij}\pa_i\psi \pa_j \chi  \biggr]d^3x, 
\label{eqdiff2}
\eeqn
where we use $S_k^{~k}=j_k u_l \tilde \gamma^{kl}/(\alpha u^t) +3P$, 
$v^k = u^k/u^t$\footnote{The above definition for $v^k$ is used only 
in this appendix.  Note that $v^k$ was differently used for
spatial velocity vector in co-moving frame with $k^\alpha$
as defined in Eq.(\ref{eq:udec}) in main sections.}, 
$\gamma^{kl} u_l = u^t (v^k+\beta^k)$ and 
Eq. (\ref{eqid1}). In the following we show that the relation 
$\Mchi=\Madm$ is equivalent to the virial relation. 

To derive the virial relation, 
we first write the general relativistic Euler equation 
$\gamma^{\nu}_{~k} \nabla_{\mu} T^{\mu}_{~\nu}=0$ in the form 
\beqn
\pa_t (j_k \sqrt{\gamma})+\pa_k (j_l \sqrt{\gamma} v^l) 
+ \pa_k (\alpha \sqrt{\gamma} P) 
+ \rho_H\psi^5 \pa_k \chi -(\rho_H+ 2S_l^{~l})\chi \psi^4 \pa_k \psi 
- \sqrt{\gamma} j_l \pa_k \beta^l
+{1 \over 2} \chi\psi S_{ij}\pa_k \tilde \gamma^{ij}=0, \label{eqEuler}
\eeqn
where $\tilde \gamma^{ij}=\gamma^{ij}\psi^{4}$. 
Equation (\ref{eqEuler}) is a fully general relativistic 
expression. In the IWM spacetimes, 
$\tilde \gamma^{ij}=\tilde \gamma_{ij}=\delta_{ij}$ and 
consequently, the last term in Eq. (\ref{eqEuler}) is neglected. 

In the following calculation, we choose the $x^1$-axis so
that, on some time-slice $\Sigma_t$, it lies along the centers of the two
members of the binary system. 

As in the Newtonian case, the virial relation can be 
derived by taking inner product with $\hat x^k$ and by performing 
an integral over three space, i.e., 
\beqn
\int d^3 \hat x \hat x^k \biggl[
\pa_t (j_k \sqrt{\gamma})+\pa_j (j_k \sqrt{\gamma} v^j) 
+ \pa_k (\alpha \sqrt{\gamma} P) 
+ \rho_H\psi^5 \pa_k \chi -(\rho_H+ 2S_l^{~l})\chi \psi^4 \pa_k \psi
- \sqrt{\gamma} j_l \pa_k \beta^l \biggr]=0. 
\eeqn
Below, we shall carry out integrals separately. 
For simplicity, we omit hats ($~\hat{}~$) on indices in the following. 

(1) First term: 
Since we assume the existence of the helical Killing vector, 
we have a relation
\beq
\pa_t {\cal J}_k =-\Omega[\pa_l(\phi^l {\cal J}_k)+{\cal J}_l \pa_k \phi^l],
\eeq
where we use $\pa_l \phi^l = 0$ and ${\cal J}_k \equiv j_k \sqrt{\gamma}$. 
In the present coordinates, $\phi^l=(-x^2, x^1-b, 0)$. 
After an integration by parts, we obtain
\beqn
\int x^k \pa_t {\cal J}_k d^3x 
=\Omega \int (\phi^k {\cal J}_k - {\cal J}_l x^k \pa_k \phi^l) d^3x. 
\eeqn
{}From a relation $x^k \pa_k \phi^l = \phi^l + \delta^{2l} b$, 
we immediately find 
\beqn
\int x^k \pa_t (j_k \sqrt{\gamma}) d^3x 
=-b \Omega \int j_2 \sqrt{\gamma} d^3x 
\equiv -b \Omega P_{\rm NS},
\eeqn
where $P_{\rm NS}$ is interpreted as 
the linear momentum of a neutron star [see Eq. (\ref{linear})]. 

(2) Second and third terms: An integration by parts immediately yields 
\beqn
&&
\int x^k \pa_j (j_k v^j \sqrt{\gamma})d^3x=-\int j_k v^k \sqrt{\gamma} d^3x,\\
&&\int x^k \pa_k (\alpha \sqrt{\gamma} P)d^3x =-3 \int \alpha \sqrt{\gamma} 
P d^3x.
\eeqn

(3) Fourth and fifth terms: 
Using Eqs. (\ref{eqpsi}) and (\ref{eqchi}), 
we can rewrite these terms as 
\beqn
\rho_H\psi^5\pa_k \chi-(\rho_H+ 2S_l^{~l})\chi \psi^4 \pa_k \psi 
=-{1 \over 2\pi} \biggl[ \Delta \psi \pa_k \chi + 
\Delta \chi \pa_k \psi \biggr] 
-{\psi^{12}K_i^{~j} K_j^{~i} \over 16\pi}\pa_k \biggl(
{\alpha \over \sqrt{\gamma}}\biggr). 
\eeqn
Taking into account an identity, 
\beqn
\int [(x^k \pa_k \psi)\Delta \chi +(x^k \pa_k \chi)\Delta \psi]d^3x 
=\int \delta^{ij} \pa_i \chi \pa_j \psi d^3 x, 
\eeqn
we find 
\beqn
&&\int x^k [\rho_H\psi^5\pa_k \chi-(\rho_H+ 2S_l^{~l})\chi \psi^4 \pa_k \psi]
d^3x \nonumber \\
&&=-\int \biggl[{1 \over 2\pi} \delta^{ij} \pa_i \chi \pa_j \psi
+ {\psi^{12}K_i^{~j} K_j^{~i} \over 16\pi}x^k \pa_k \biggl(
{\alpha \over \sqrt{\gamma}}\biggr)\biggr]d^3x.
\eeqn

(4) Sixth term: 
\beqn
&& -\int \sqrt{\gamma} j_i x^k \pa_k \beta^i d^3x 
=-{1 \over 8\pi} \int \pa_j (\sqrt{\gamma} K_i^{~j}) x^k\pa_k \beta^i d^3x 
\nonumber \\
&&={1 \over 8\pi} \int [\sqrt{\gamma} K_i^{~j} x^k \pa_k\pa_j \beta^i 
+\sqrt{\gamma} K_i^{~j} \pa_j \beta^i]d^3x 
-{1 \over 8\pi} \oint
\sqrt{\gamma} K_i^{~j} x^k \pa_k \beta^i \ds_j  \nonumber \\
&&=-{1 \over 8\pi} \int [\pa_k(\sqrt{\gamma} K_i^{~j}) x^k \pa_j \beta^i 
+2 \sqrt{\gamma} K_i^{~j} \pa_j \beta^i]d^3x 
+{1 \over 8\pi} \oint
\sqrt{\gamma} (K_i^{~j} x^k-K_i^{~k} x^j) \pa_j \beta^i \ds_k  \nonumber \\
&&=-{1 \over 8\pi} \int \biggl[{\alpha x^k \over 2 \sqrt{\gamma}}
\pa_k(\psi^{12} K_i^{~j} K_j^{~i})
+2 \alpha \sqrt{\gamma} K_i^{~j} K_j^{~i} \biggr]d^3x 
+{1 \over 8\pi} \oint
\sqrt{\gamma} (K_i^{~j} x^k-K_i^{~k} x^j) \pa_j \beta^i \ds_k  \nonumber \\
&&=-{1 \over 16\pi} \int \biggl[
\alpha \sqrt{\gamma} K_i^{~j} K_j^{~i} 
-\psi^{12} K_i^{~j} K_j^{~i} x^k\pa_k \biggl(
{\alpha \over \sqrt{\gamma}}\biggr)\biggr]d^3x \nonumber \\
&&~~~-{1 \over 16\pi}\oint \alpha \psi^{6} K_i^{~j} K_j^{~i} x^k \ds_k
+{1 \over 8\pi} \oint
\sqrt{\gamma} (K_i^{~j} x^k-K_i^{~k} x^j) \pa_j \beta^i \ds_k  \nonumber \\
&&={1 \over 16\pi} \int \biggl[
8\pi \sqrt{\gamma} j_k \beta^k +\psi^{12} K_i^{~j} K_j^{~i} x^k\pa_k \biggl(
{\alpha \over \sqrt{\gamma}}\biggr)\biggr]d^3x \nonumber \\
&&~~~+{1 \over 16\pi}\oint \alpha \psi^{6} K_i^{~j} K_j^{~i} x^k \ds_k
-{1 \over 8\pi} \oint \sqrt{\gamma} K_i^{~k} x^j \pa_j \beta^i \ds_k,
\label{eq66}
\eeqn
where we use Eqs. (\ref{momcon}) and (\ref{eqid1}). 
Because of the asymptotic behavior,  
the surface terms at $r\rightarrow \infty$ 
in the last line of Eq. (\ref{eq66}) vanish. 
Eq. (\ref{linear}) implies 
that the center of mass of the system does not move in the $x^2$ direction,
that the sum of the momenta of the neutron stars vanishes. 

Gathering the results of (1)--(6), 
we obtain the relation
\beqn
0 = -\int \biggl(j_k v^k \sqrt{\gamma} + 3 \alpha \sqrt{\gamma} P 
+{1 \over 2\pi}\delta^{ij}\pa_i \chi \pa_j \psi
-{1 \over 2} \sqrt{\gamma} j_k \beta^k \biggr) d^3x.
\label{virial}
\eeqn
This is the virial relation for a neutron star binary system in
quasiequilibrium.

{}From Eq. (\ref{eqdiff2}), the right-hand side of Eq. (\ref{virial}) 
is written as
\beqn
0=-{\Mchi - \Madm \over 2},
\eeqn
implying $\Madm=\Mchi=M_K$, if the virial relation holds.

%%%%%%%%%%%%%

\section{The first law for Newtonian binary systems }
\label{appendix3}

In this appendix, we derive a first law of thermodynamics for Newtonian 
gravity.  We start with a first-order perturbation of the energy of 
a perfect-fluid. 
\beqn
	E = T\,+\,W\,+\,U, 
\eeqn
where
\beq
	T = \int_V\frac{1}{2} \rho\, v^2 dV, \qquad
	W = \int_V \left(\rho\, \Phi_{\rm N}\,
	+\,\frac{1}{8\pi G}\na_i \Phi_{\rm N}\na_i \Phi_{\rm N}\right)\,dV, 
\qquad
	U = \int_V \rho\, u\, dV.
\eeq
and $\Phi_{\rm N}$ and $u$ denote the Newtonian potential and 
specific internal energy\footnote{We use $v^i$ for the fluid velocity vector 
in the inertial frame in this appendix.  Note that it was differently used for 
spatial velocity vector in co-moving frame with $k^\alpha$ 
as defined in Eq.(\ref{eq:udec}) in main sections.}. 
An integral equation 
\beq
\label{eq:AintfN}
	\dl \int_V f\, \rho\, dV = 
	\int_V \Dl f \, \rho\, dV + 
	\int_V f\, \Dl(\rho\, dV), 
\eeq
is satisfied for a perturbation.  

The perturbation of the kinetic energy $T$ can be expressed as 
follows
\beq
\label{eq:AdTN}
	\dl T = 
	\int_V \rho v^i \Dl v_i 
	+ \int_V \frac{1}{2}\, v^2 \Dl(\rho\, dV)
	+ \int_V\left[\xi^j v_j\na_i (\rho v^i) 
	+ \rho\, \xi^j v^i \na_i v_j \right]dV
	- \oint_{\pa V} \rho\, v^i v_j \xi^j dS_i
\eeq
where we used the relation,  
\beq
	\frac{1}{2} \Dl v^2 =  v^i \Dl v_i - v^i v_j \na _i \xi^j.  
\eeq
The perturbation of the gravitational potential energy 
becomes 
\beq
\label{eq:AdWN}
	\dl W = 
	\int_V  \Phi_{\rm N} \Dl(\rho\, dV) 
	+ \int_V \rho\,\xi^i \na_i \Phi_{\rm N}\, dV
	- \frac{1}{4\pi G}\int_V (\na^2\Phi_{\rm N} - 4\pi G\rho)\,\dl
           \Phi_{\rm N}\, dV
	+ \frac{1}{4\pi G} \oint_{\pa V} \na_i \Phi_{\rm N}\,\dl\Phi_{\rm N}\,
          dS_i. 
\eeq
The perturbation of the internal energy becomes 
\beq
\label{eq:AdUN}
	\dl U =
	\int_V  \rho\, T\Dl s\,dV 
	+ \int_V  \left(u + \frac{P}{\rho}\right)\Dl(\rho\, dV)
	+ \int_V \xi^i\na_i P\,dV - \oint_{\pa V} \xi^i\,P\,dS_i , 
\eeq
where we used a relation 
\beq
\label{eq:Adv}
	\na_i \xi^i  dV  = \Dl(dV), 
\eeq
as well as a local thermodynamic relation, 
\beq
	\Dl u = T \Dl s + \frac{P}{\rho^2}\Dl\rho. 
\eeq
Surface integrals appeared in expressions for $\dl T$, $\dl W$ and $\dl U$
are all vanish.  
Combining Eqs.(\ref{eq:AdTN}), (\ref{eq:AdWN}) and 
(\ref{eq:AdUN}), we have a perturbation of the Newtonian energy integral:  
\beqn
\label{eq:AdEN}
	\dl E &=& \dl T + \dl W + \dl U 
\nonumber \\
	&=& \int_V  \rho\, T\Dl s\,dV 
	+ \int_V \left(\frac{1}{2}\, v^2 +\Phi_{\rm N} + 
	u + \frac{P}{\rho}\right) \Dl(\rho\, dV) 
	+ \int_V \rho v^i \Dl v_i \,dV
\nonumber \\
	&&+ \int_V\left[\xi^j v_j\na_i (\rho v^i) 
	+ \xi^i \rho \left(v^j \na_j v_i + \frac{1}{\rho}\na_i P 
	+ \na_i \Phi_{\rm N}\, \right)
	- \frac{1}{4\pi G}(\na^2\Phi_{\rm N} - 4\pi G\rho)\,\dl\Phi_{\rm N}\,\right]dV
\nonumber \\
	&=& \int_V  \rho\, T\Dl s\,dV 
	+ \int_V \left(\frac{1}{2}\, v^2 +\Phi_{\rm N} + 
	u + \frac{P}{\rho}\right) \Dl(\rho\, dV) 
	+ \int_V \rho v^i \Dl v_i \,dV 
\nonumber \\
	&&+ \int_V \xi^j v_j\left[\frac{\pa\rho}{\pa\, t} 
	+ \na_i (\rho v^i)\right] dV
	+ \int_V \xi^i \rho \left(\frac{\pa v_i}{\pa\, t} + v^j \na_j v_i 
	+ \frac{1}{\rho}\na_i P + \na_i \Phi_{\rm N}\, \right)dV
\nonumber \\
	&&- \frac{1}{4\pi G}\int_V(\na^2\Phi_{\rm N} - 4\pi G\rho)\,\dl\Phi_{\rm N}\,dV
	- \int_V \xi^i \frac{\pa\rho v_i}{\pa\, t} dV.
\eeqn

Next, we derive a variation of the total angular momentum $J$ 
defined by 
\beq
	J = \int_V \rho\,v_i\phi^i\, dV ,
\eeq
where $\phi^i$ is a generator of rotation with Cartesian components  
$\phi^i=(-y,x,0)$.  The variation of $J$ is 
\beq
\label{eq:AdJ1}
	\dl J = 
	\int_V \rho\,\Dl v_i\,\phi^i\, dV 
	+ \int_V \rho\, v_i\,\Dl \phi^i\, dV 
	+ \int_V v_i\,\phi^i\,\Dl(\rho\, dV).  
\eeq
Using a relation 
\beq
	\Dl\phi^i = \dl\phi^i + \Lie_\xi \phi^i 
	= - \Lie_\phi \xi^i, 
\eeq
the second term of Eq.(\ref{eq:AdJ1}) is rewritten as follows:
\beqn
	\int_V \rho\, v_i\,\Dl \phi^i\, dV 
	&=& - \int_V \rho\, v_i\,\Lie_\phi \xi^i\, dV 
	= - \int_V \Lie_\phi(\rho\, v_i\,\xi^i)\, dV 
	+ \int_V \xi^i\,\Lie_\phi(\rho\, v_i)\, dV 
\nonumber \\
	&=& - \oint_{\pa V} \rho\, v_i\,\xi^i \phi^j\, dS_j 
	+ \int_V \xi^i\,\Lie_\phi(\rho\, v_i)\, dV , 
\eeqn
where we used $\na_j \phi^j = 0$.  Discarding the surface term 
in the above expression and substituting in  Eq.(\ref{eq:AdJ1}), we have
a variation of the total angular momentum $\dl J$ as follows:
\beq
\label{eq:AdJN}
	\dl J = 
	\int_V v_i\,\phi^i\,\Dl(\rho\, dV)  
	+ \int_V \rho\,\phi^i\,\Dl v_i\, dV 
	+ \int_V \xi^i\,\Lie_\phi(\rho\, v_i)\, dV , 
\eeq

Finally we write down a general expression for the combination of 
$\dl E$ and $\Omega \dl J$, where $\Omega$ is a constant parameter, 
\beqn
\label{eq:AdQN}
	\dl E &-& \Omega \dl J = 
\nonumber \\
	&& \int_V  \rho\, T\Dl s\,dV 
	+ \int_V \left(\frac{1}{2}\, v^2 +\Phi_{\rm N} + u + \frac{P}{\rho}
	- v_i\,\Omega\phi^i\right) \Dl(\rho\, dV) 
	+ \int_V \rho (v^i-\Omega\,\phi^i)\,\Dl v_i \,dV
\nonumber \\
	&+& \int_V \xi^j v_j\left[\frac{\pa\,\rho}{\pa\, t} 
	+ \na_i (\rho v^i)\right] dV
	+ \int_V \xi^i \rho \left(\frac{\pa\, v_i}{\pa\, t} + v^j \na_j v_i 
	+ \frac{1}{\rho}\na_i P + \na_i \Phi_{\rm N}\, \right)dV
	- \frac{1}{4\pi G}\int_V(\na^2\Phi_{\rm N} - 4\pi G\rho)\,\dl\Phi_{\rm N}\,dV
\nonumber \\
	&-& \int_V \xi^i \left[\frac{\pa\,\rho v_i}{\pa\, t} 
	+ \Lie_{\Omega\phi}(\rho\, v_i)\right]\, dV , 
\nonumber \\
\eeqn

As an application of the above general expression,  consider a Newtonian 
binary star system in circular orbit\cite{ue98}.  
In this case, the fluid variables 
${\cal Q}$ admit a helical symmetry, namely,  
\beq
	\left[\frac{\pa}{\pa\, t} + \Omega\Lie_{\phi}\right]{\cal Q} = 0, 
\eeq
that is, the last integral in Eq.(\ref{eq:AdQN}) vanishes.  
When a Mass conservation equation, the Euler equation and the Poisson 
equation for the Newtonian gravity are satisfied, namely, 
\beq
\label{eq:AeqN}
	\frac{\pa\,\rho}{\pa\, t} + \na_i (\rho v^i) = 0, \qquad
	\frac{\pa\, v_i}{\pa\, t} + v^j \na_j v_i 
	= - \frac{1}{\rho}\na_i P - \na_i \Phi_{\rm N}, \quad{\rm and}\quad
	\na^2\Phi_{\rm N} = 4\pi G\rho, 
\eeq
Eq.(\ref{eq:AdQN}) takes a simpler form,   
\beq
\label{eq:A1st}
	\dl E = \Omega \dl J 
	+ \int_V  \rho\, T\Dl s\,dV
	+ \int_V \left(\frac{1}{2}\, v^2 +\Phi_{\rm N} + u + \frac{P}{\rho}
	- v_i\,\Omega\phi^i\right) \Dl(\rho\, dV) 
	+ \int_V \rho (v^i-\Omega\,\phi^i)\,\Dl v_i \,dV.  
\eeq
If we further assume that the perturbed flow is isentropic and 
mass conserving, 
\beq
\Dl s = 0 \quad{\rm and}\quad \Dl(\rho\, dV) = 0
\eeq
and that the vorticity of each fluid elements is conserved, 
\beq
\label{eq:AvoN}
\Dl\,\omega_{ij} = \Dl(\na_j v_i - \na_i v_j) 
= \na_j \Dl v_i - \na_i \Dl v_j = 0, 
\eeq
then Eq.(\ref{eq:A1st}) reduces to 
\beq
\dl E = \Omega\, \dl J. 
\eeq
Here we have used Eq.(\ref{eq:AvoN}) to introduce a function 
$\Psi$ for which 
\beq
\na_i \Psi = \Dl v_i; 
\eeq
this form of $\Delta v_i$, together with helical symmetry imply that 
the last term in Eq.(\ref{eq:A1st}) vanishes:
\beq
\int_V \rho (v^i-\Omega\,\phi^i)\,\Dl v_i \,dV
= \int_V \rho (v^i-\Omega\,\phi^i)\,\na_i \Psi \,dV
= \oint_{\pa V} \rho (v^i-\Omega\,\phi^i)\, \Psi \,dS_i
+ \int_V \left(\frac{\pa\,\rho}{\pa\, t} 
+ \Lie_{\Omega\,\phi}\rho \right)\, \Psi \,dV = 0
\eeq
where we used mass conservation equation and $\na_i \phi^i = 0$. 

\end{document}